\documentclass[final,5p,times,twocolumn,authoryear]{elsarticle}
\usepackage[utf8]{inputenc}
\usepackage[colorlinks=true,linkcolor=black, citecolor=blue, urlcolor=blue]{hyperref}
 
\usepackage{subfig, amsmath, upgreek, siunitx, fancyhdr}
\usepackage[export]{adjustbox}

\usepackage[colorinlistoftodos]{todonotes}

\usepackage{amssymb}
\journal{Astroparticle Physics}

\fancypagestyle{cclicense}{
\lhead{\copyright 2021. This manuscript version is made available under the CC-BY-NC-ND 4.0 license http://creativecommons.org/licenses/by-nc-nd/4.0/}
}

\begin{document}

\begin{frontmatter}



\title{Cadmium Zinc Telluride Detectors for a Next-Generation Hard X-ray Telescope}
\author[WUSTL]{J. Tang \corref{cor}}
\author[UNH]{F. Kislat}
\author[WUSTL]{H.~Krawczynski}

\address[WUSTL]{Washington University in St. Louis, Physics Department, 1 Brookings Dr., CB 1105, St. Louis, MO 63130}
\address[UNH]{University of New Hampshire, Department of Physics \& Astronomy and Space Science Center, 8 College Rd, Durham, NH 03824}
\cortext[cor]{Corresponding author}
%\address[BNL]{Brookhaven National Laboratory, 98 Rochester St, Upton, NY 11973}

\begin{abstract}
We are currently developing Cadmium Zinc Telluride (CZT) detectors for a next-generation space-borne hard X-ray telescope which can follow up on the highly successful {\it NuSTAR} (Nuclear Spectroscopic Telescope Array) mission. Since the launch of {\it NuSTAR} in 2012, there have been major advances in the area of
X-ray mirrors, and state-of-the-art X-ray mirrors can improve on {\it NuSTAR's} angular resolution of
\SI{\sim1}{arcmin} Half Power Diameter (HPD) to \SI{15}{\arcsecond} or even \SI{5}{\arcsecond} HPD. Consequently, the size of the detector pixels must be reduced to match this resolution. This paper presents detailed simulations of 
relatively thin (\SI{1}{mm} thick) CZT detectors with hexagonal pixels at a next-neighbor distance of \SI{150}{\upmu m}. The simulations account for the non-negligible spatial extent of the deposition of the energy of the incident photon, and include detailed modeling of the spreading of the free charge carriers  as they move toward the detector electrodes.
We discuss methods to reconstruct the energies of the incident photons, and the locations where
the photons hit the detector. We show that the charge recorded in the brightest pixel
and six adjacent pixels suffices to obtain excellent energy and spatial resolutions.
The simulation results are being used to guide the design of a hybrid 
application-specific integrated circuit (ASIC)-CZT detector package.
\end{abstract} 

\end{frontmatter}

\thispagestyle{cclicense}

%%%%%%%%%%%%%%%%%%%%%%%%%%%%%%%%%%%%%%%%%%%%%

\section{Introduction}

%%%%%%%%%%%%%%%%%%%%%%%%%%%%%%%%%%%%%%%%%%%%%
Cadmium Zinc Telluride (CZT) detectors are an attractive detector technology for hard X-ray astronomy as they offer excellent spatial resolutions,
good energy resolutions, and, compared to Si and Ge detectors, much larger photoelectric effect cross sections at hard X-ray energies.
A CZT imager may be used on a next-generation telescope succeeding the 
space-based hard X-ray telescope {\it NuSTAR} \citep{nustar}. 
The high stoppping power and excellent energy resolution of the {\it NuSTAR}
CZT detectors enabled it to image the Cassiopeia A (Cas A) supernova remnant in the \SI{67.9}{keV} and \SI{78.4}{keV} line emissions from the radioactve isotope $^{44}$Ti \citep{2014Natur.506..339G, Boggs670, 2017ApJ...834...19G}.  Contrary to the soft X-ray lines detected previously,
the nuclear $^{44}$Ti emission directly tracks the yield of nuclear material independent of the temperature and density of the ejecta \citep{2014Natur.506..339G, Boggs670,2017ApJ...834...19G}.

The recently developed monocrystalline silicon X-ray mirrors \citep{zhangMirrors}
or electro-formed-nickel replicated (ENR) X-ray optics \citep{Gask:18} 
promise angular resolutions with Half Power Diameters (HPD) of between a few arcseconds 
and 15 arcseconds -- even at hard X-ray energies.  
The proposed {\it HEX-P} \citep{2018SPIE10699E..6MM}
and {\it BEST} \citep{krawczynski2012black} observatories 
seek to capitalize on this technology, as the point source sensitivity
scales linearly with the angular resolution. 
Nyquist sampling the images provided by the improved X-ray mirrors requires
detectors with excellent spatial resolutions. Our group is thus 
leading the development of new small-pixel CZT detectors 
with center-to-center pitch of 150 microns and hexagonal pixels,
improving by a factor of four over {\it NuSTAR's} CZT detectors (605-micron pixel pitch).

This paper discusses the simulations performed for the design of the third-generation  Hyperspectral Energy-resolving X-ray Imaging Detector \citep[HEXID3][]{Li:17,Li:18}, which features hexagonal pixels 
at a next-neighbor pitch of \SI{\sim 150}{\upmu m} and uses a low noise 
front end design achieving a projected readout noise of 14 electrons
Root Mean Square (RMS).  
The advantage of using hexagonal over square pixels is that all the nearest neighbors of any given pixel are equivalent; in square pixels, some immediate neighbors are closer than others.
Another similar ASIC for hybridization with pixelated CZT detectors is the
High Energy X-ray Imaging Technology (HEXITEC) ASIC developed by 
Rutherford Appleton Laboratory. The HEXITEC ASIC features 6400 square 
pixels at a next-neighbor pitch of \SI{250}{\upmu m} with an electronic 
readout noise of 50 electrons RMS \citep{2016SPD....47.0811R,2016SPIE.9915E..1DB,2017SPIE10397E..02R}.

Our simulations model in detail the interactions of the incident photons, secondary photons and
high-energy electrons generated in the CZT, and the ionization losses of the latter.
The simulations furthermore model the drift and diffusion of the negative and positive charge
carriers through the CZT, including the effects of mutual repulsion of charge carriers
of equal polarity. This detailed treatment allows us to predict the 
properties of the signals, including the pixel multiplicity, and the dependence of the 
pixel signals on where in the detector the free charge carriers are generated.
Earlier discussions of CZT detector simulations can be found in
\citep{BENOIT2009508,2011SPIE.8145E..07K,2017SPIE10397E..02R,2019SPIE10948E..1VP,2019SPIE10948E..4GL}.
Compared to the earlier study of small pixel detectors of \citep{2017SPIE10397E..02R}, 
the shape of our charge clouds evolve owing to charge carrier repulsion and diffusion as the clouds drift inside the detector. Furthermore, we extend the study from a pixel pitch of \SI{250}{\upmu m} to smaller \SI{150}{\upmu m} pixels.

The rest of the paper is organized as follows. After describing the 
detector simulation methodology in Section \ref{method}, we present the
results of the simulations in Section \ref{results}. 
Our studies show that the \SI{1}{mm} thick 
detectors have a limited energy range over which they 
give excellent performance.  We discuss the results and implications 
for the camera of a {\it NuSTAR} follow-up mission in Section \ref{disc}.

%%%%%%%%%%%%%%%%%%%%%%%%%%%%%%%%%%%%%%%%%%

\section{Simulations of the CZT/ASIC Hybrid Detectors} \label{method}

%%%%%%%%%%%%%%%%%%%%%%%%%%%%%%%%%%%%%%%%%%

\begin{figure}
    \centering
    \includegraphics[width=\linewidth]{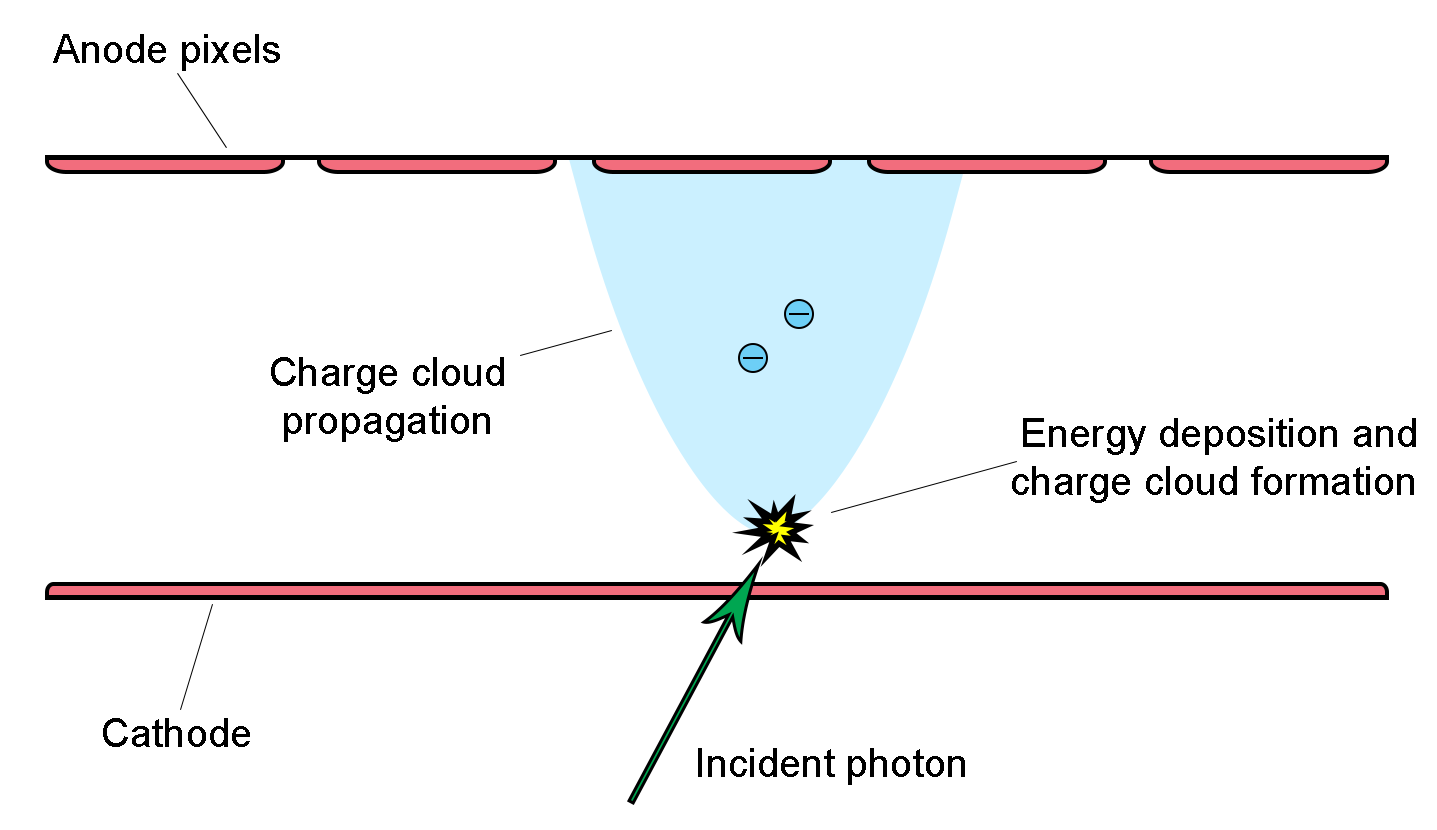}
    \caption{Sketch showing how an energy deposition leads to the generation of a free electron cloud that 
    follows the applied electric field to the anode contacts, and widens owing to the effects of 
    diffusion and repulsion of charges of equal polarity. Our model captures the effects of charge carriers being created along 
    one or sometimes several photo-electron tracks, and accounts not only for drifting electrons (shown here), 
    but also for drifting holes (not shown).}
    \label{fig:detector}
\end{figure}

X-rays impinging on a CZT detector interact via photoelectric, scattering, and pair production interactions. 
Photoelectric interactions dominate up to primary photon energies of $E_{\gamma}<\SI{240}{keV}$ at which Compton scattering becomes dominant (assuming 40\% Cd, 10\% Zn and 50\% Te).
The photo-electron of a photoelectric effect interaction loses most of its energy to ionization. 
The ionization promotes electrons to the conduction band, generating clouds of electrons and holes.
Applying a bias across the detector causes the charge carriers to drift and to induce the read-out signal. 
While drifting, the charge clouds expand due to the 
repulsion and attraction between charge carriers and diffuse owing to
spatial concentration gradients. 
In small-pixel detectors, the charge clouds can spread and induce currents over multiple pixels, 
resulting in charge sharing (Fig. \ref{fig:detector}). 

The simulations consist of three parts: simulations of the 
photon interaction and subsequent energy losses of the 
photo-electron, Compton electron, or produced pairs and their
secondaries, a simulation of the detector drift field, and charge cloud tracking and signal integration. 
\subsection{{\tt Geant4} Simulations}\label{geant}
The interaction of the photons with the \SI{1}{mm} thick detector are simulated with the {\tt Geant4} simulation toolkit \citep{geant} at discrete photon energies.
% We use a random number generator to throw the simulated events uniformly over a unit rectangular cell centered on one pixel. 
The resulting interactions and energy depositions are 
stored for subsequent post-processing. 

\begin{figure}
    \centering
    \includegraphics[width=0.95\columnwidth]{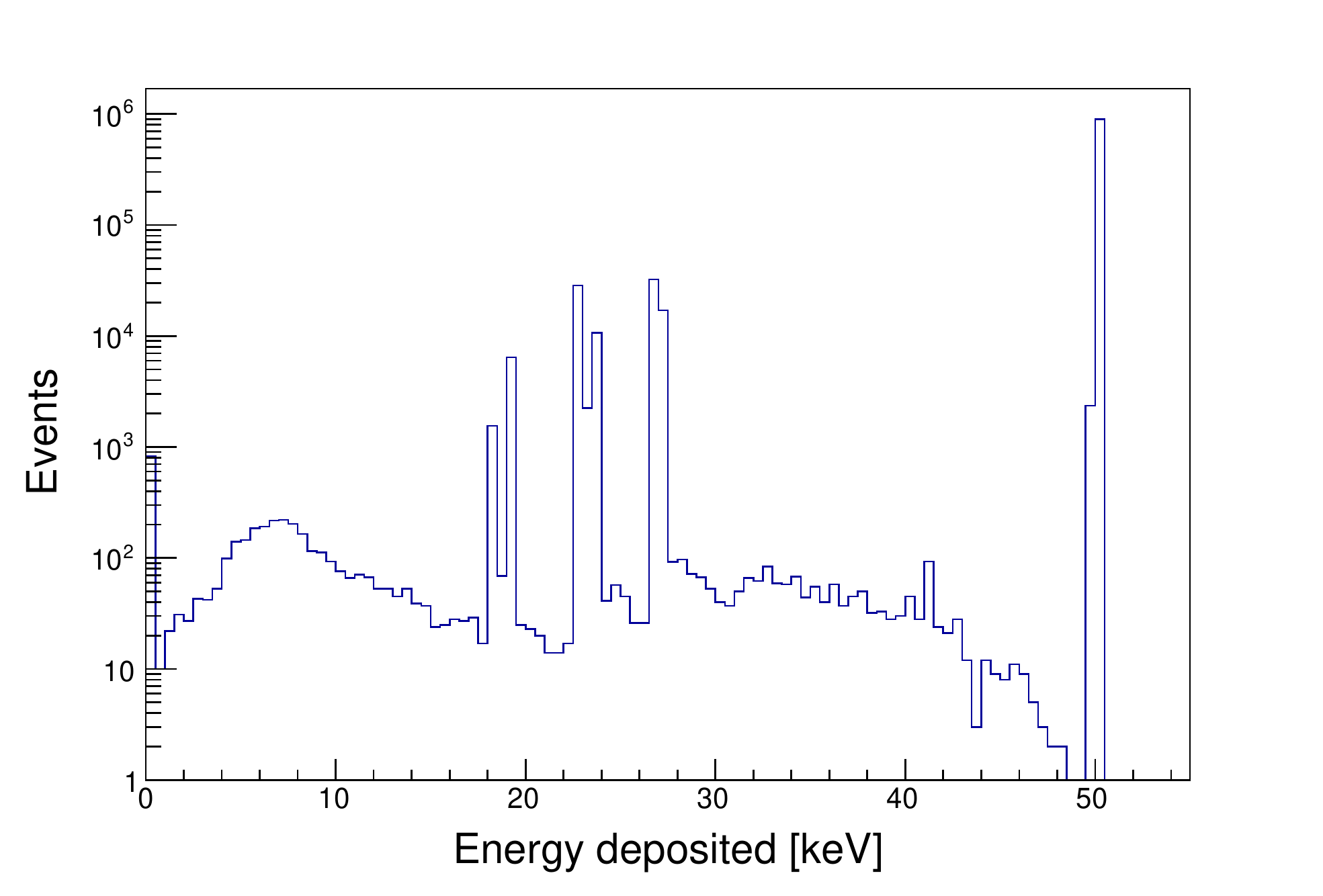}
    \includegraphics[width=0.95\columnwidth]{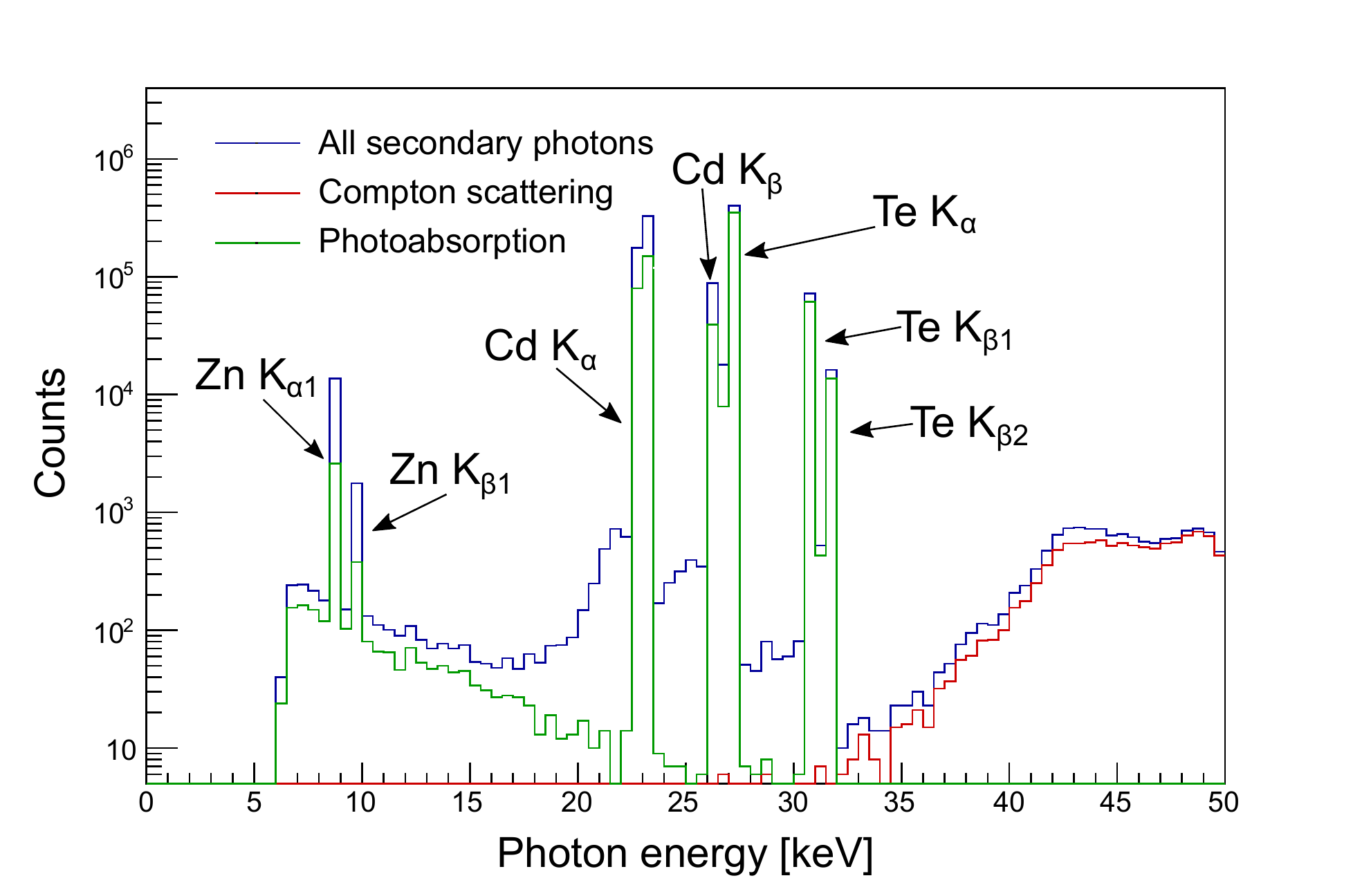}

    \caption{
    \emph{Top:} Distribution of the total energy deposited by 50 keV photons in the CZT detectors.
    In this simulation of $10^6$ events, 89.3\% have the full \SI{50}{keV} deposited.
    For the remaining events, some energy escapes the detector, leading to substantially smaller total energy deposits.
    \emph{Bottom:} For the same incident 50 keV photons as shown in the upper panel, the lower panel 
    presents the distribution of the energies of photons produced in the CZT detector, either 
    through Compton scattering, or, following photoelectric absorption and the inelastic scattering of the
    photo-electron, through the emission of fluorescent photons. Prominent fluorescence lines are labeled. 
    }
    \label{fig:geantStuff}
\end{figure}

The upper panel of Figure \ref{fig:geantStuff} shows the energies deposited in the simulated CZT detector 
by a beam of 50 keV photons. A large fraction of 89.3\% of the photons deposit their full energy in the detector. 
Some prominent escape peaks can be recognized.  The peaks correspond to the initial energy of \SI{50}{keV} minus the fluorescence photon energies. Secondary photons from an earlier interaction may escape the detector, resulting in incomplete energy detection, as shown by the energy depositions less than \SI{50}{keV}.
The lower panel of Figure \ref{fig:geantStuff} shows the energy of the photons produced in the CZT and identifies several prominent X-ray lines of the Cd, Zn, and Te atoms. 
The objective of the optimization of the detector design is to reconstruct the deposited energy as well as possible. The energy escaping the detector can, of course, not be recovered, and thus
contributes along the Fano factor limit to the theoretical best possible achievable 
energy resolution. 
  
\subsection{Potential Simulations}\label{pot}
The electrostatic potential $\phi$ and electric field inside the detector are inferred from solving Gauss' law with Dirichlet boundary conditions:
\begin{equation}
	\nabla \cdot \mathbf{D} = -\nabla\epsilon \cdot \nabla\phi - \epsilon \nabla^2\phi = 0.
    \label{eq:Gauss}
\end{equation}
Here, {\bf D} is the electric displacement field, and $\epsilon\approx10\epsilon_0$ is the electric permittivity of CZT  \citep{2011SPIE.8145E..07K}. 
Eq. \eqref{eq:Gauss} is solved on a three-dimensional discrete rectangular grid. The detector has a monolithic cathode in the $z = 0$ plane and a hexagonally-pixelated anode at the $z=d$ plane, where $d$ is the thickness of the detector. 
The $x$ and $y$ directions are equivalent in terms of the grid resolution needed, so we set $\delta x = \delta y \not= \delta z$, where $\delta x$, $\delta y$, and $\delta z$ denote the grid spacing in the $x$, $y$, and $z$ directions, respectively. We expect the potential to closely resemble that of a parallel plate capacitor throughout most of the detector, so we take $\delta z$ to be relatively large compared to $\delta x$ and $\delta y$.
We use the finite difference approximations
\begin{equation}
    \frac{df}{dx} = \frac{f(x+h) - f(x-h)}{2h} + O(h^2), 
    \label{eq:firstDerivative}
\end{equation}
\begin{equation}
  \frac{d^2f}{dx^2} =\frac{f(x+h) - 2f(x) + f(x-h)}{h^2} + O(h^2)
  \label{eq:secondDerivative}
\end{equation}
to solve Eq.\ (\ref{eq:Gauss}) with the Successive Overrelaxation Method (SOR) \citep{numerical}. In each iteration, 
the electric potential is updated to the average of the 
surrounding values, including a certain amount of ``overshooting'' to accelerate convergence: 
\begin{equation}
	\phi \mapsto \omega\phi^* + (1 - \omega)\phi,
    \label{eq:SOR}
\end{equation}
where $\omega$ is called the overrelaxation factor and $\phi^*$ is calculated by solving for the potential at the center grid point after substituting Eqs. \eqref{eq:firstDerivative} and \eqref{eq:secondDerivative} into Eq. \eqref{eq:Gauss}. 
The method converges for values of $\omega$ between 1 and 2. 
The running time can be reduced by starting with an initial guess that is close to the expected potential. 
The final potential is expected to be close to that of a parallel plate capacitor, so the starting configuration was a simple gradient in $z$.

The error is calculated using the norm of an error vector $\mathbf{e}$, where $e_i = (\nabla \cdot \mathbf{D})_i$, where $(\nabla \cdot \mathbf{D})_i$ ($\nabla \cdot \mathbf{D}$ calculated at the $i^{\text{th}}$ grid point) is calculated using Eq. \eqref{eq:Gauss} and the discrete derivatives. For a tolerance $\varepsilon$ and $G$ grid points, we consider the simulation converged if $|\mathbf{e}| < \varepsilon G$. For our simulations, we chose a tolerance of $10^{-6}$.

\begin{figure*}
    \centering
    \hspace{0.15cm}
    \includegraphics[width=0.42\linewidth]{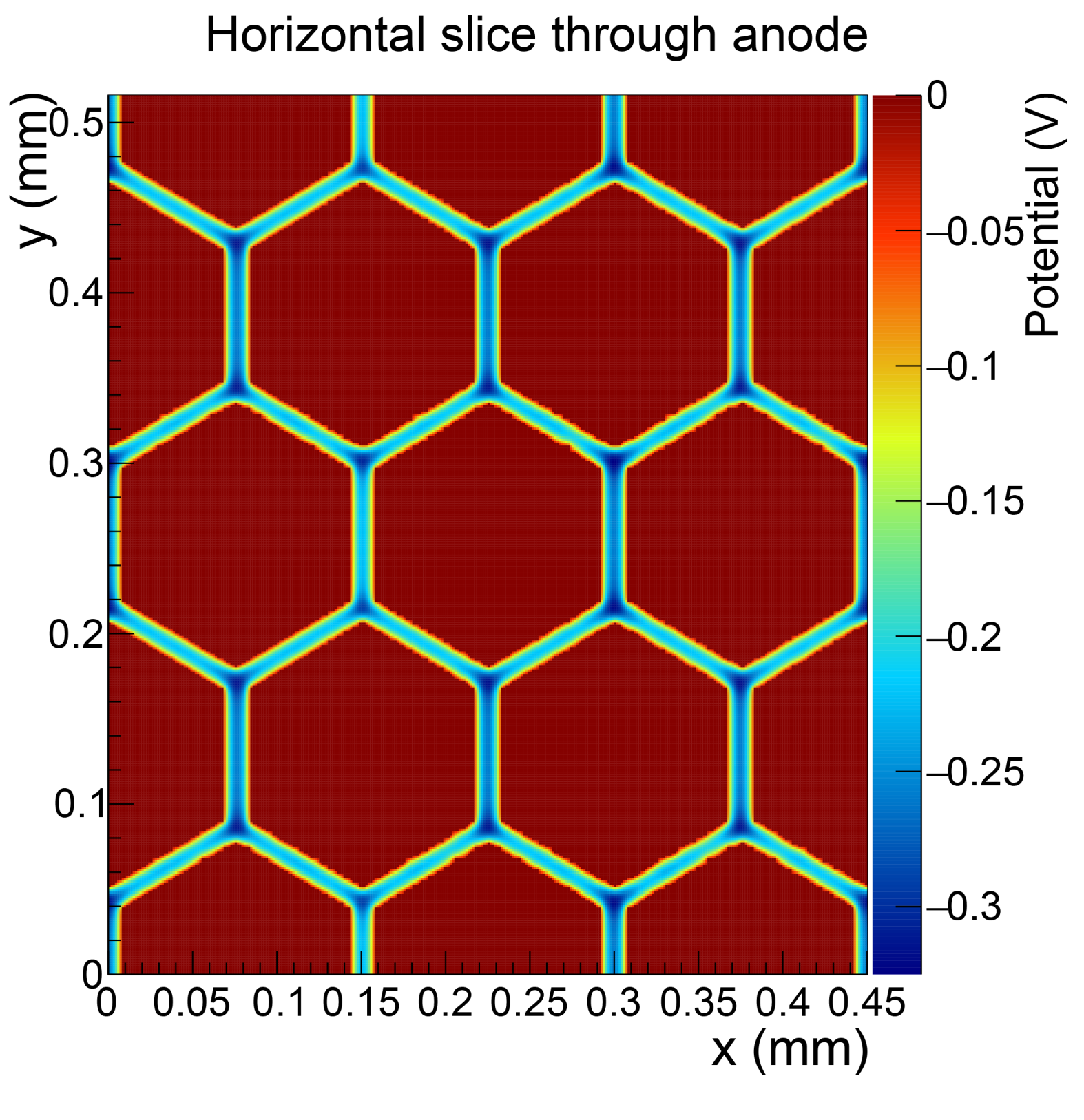}
    \hspace{0.45cm}
    \includegraphics[width=0.42\linewidth]{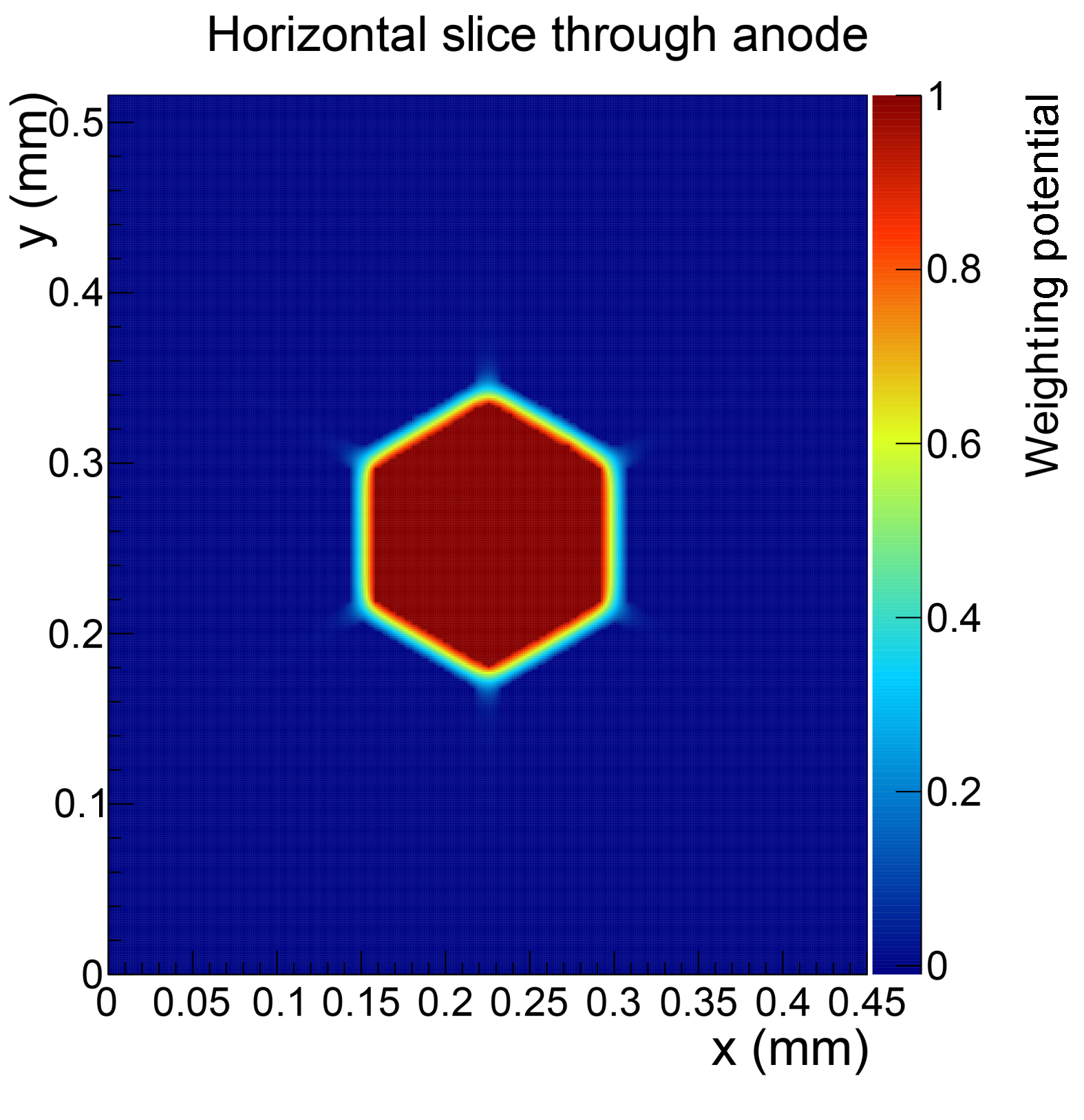}
    
    \includegraphics[width=0.4\linewidth]{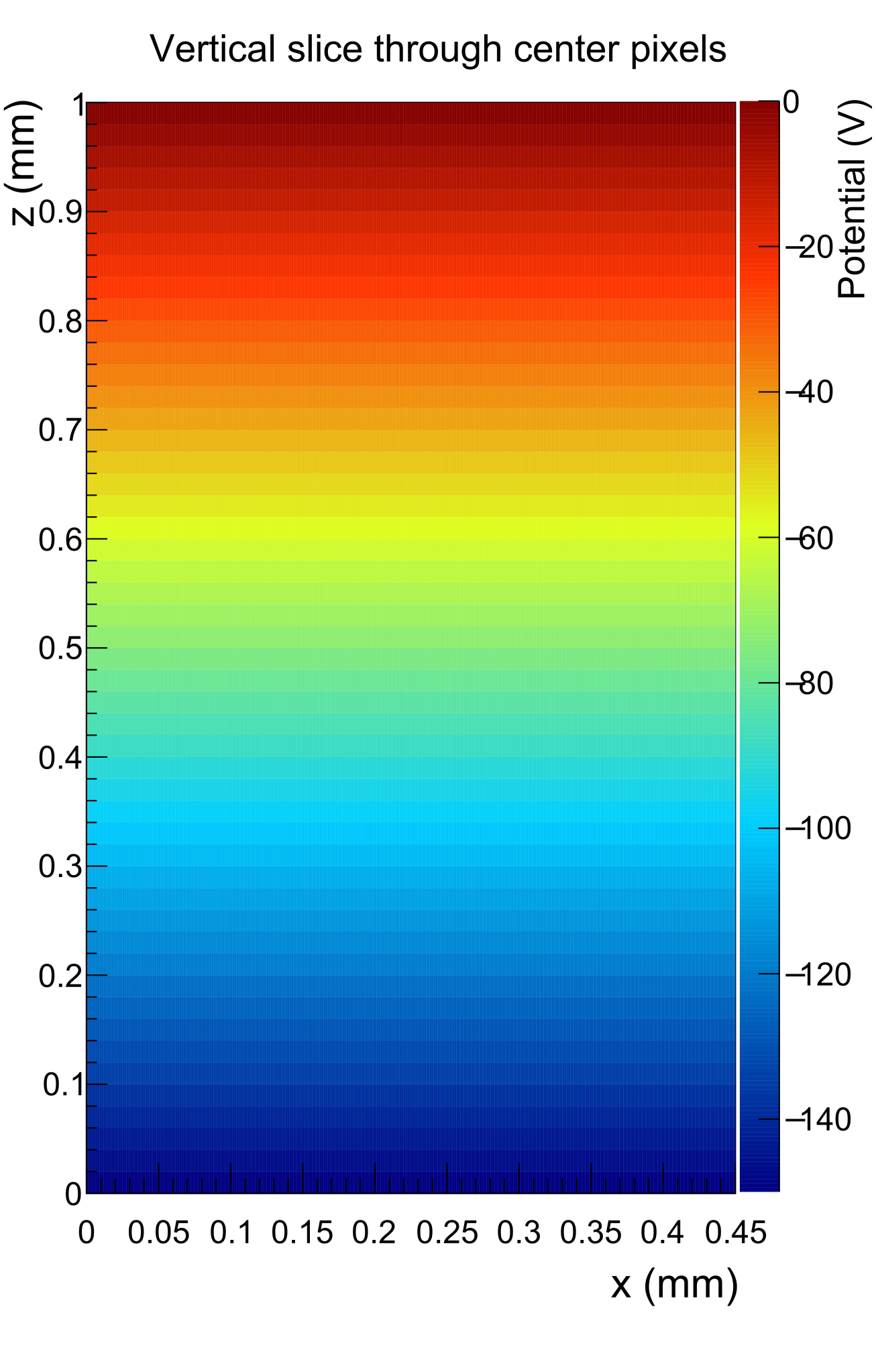}
    \hspace{0.9cm}
    \includegraphics[width=0.4\linewidth]{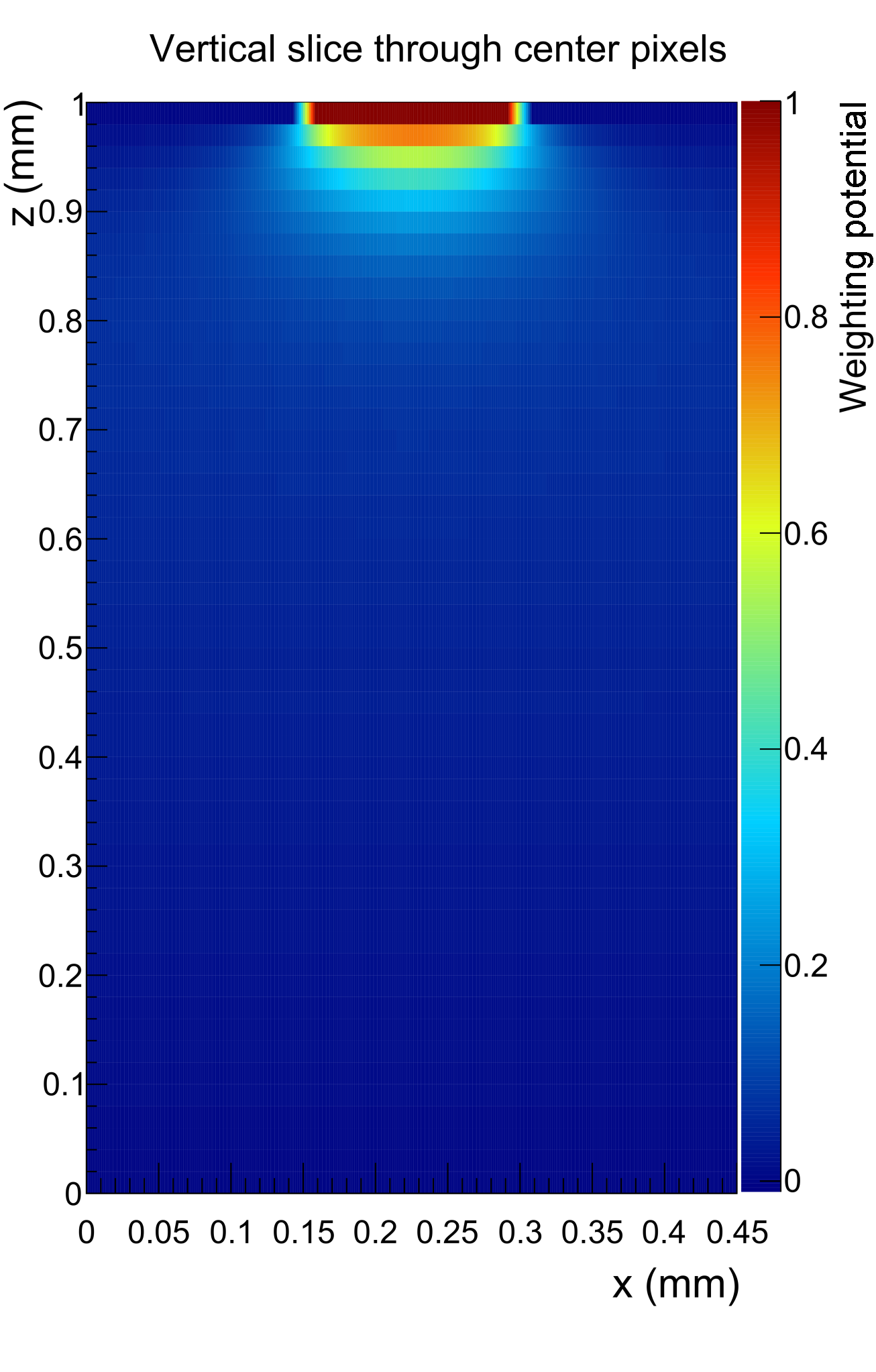}
    \caption{Two-dimensional slices of the three-dimensional simulated potentials. 
    This geometry features a \SI{1}{mm} thick detector, \SI{150}{\upmu m} center-to-center pitch with \SI{15}{\upmu m} spacing between pixels, and a \SI{-150}{V} bias. 
    The bottom slices (vertical slices through the center pixels) only show the potential inside the detector; however, there is additional vacuum simulated up to $z = \SI{2}{mm}$, at which there is a grounded plane. 
    This serves as the upper boundary condition. 
    \emph{Left}: Slices of the electric potential. 
    This is the drift field which pulls electrons toward the pixelated anode side and holes towards the cathode side of the detector.
    \emph{Right}: Slices of the weighting potential. 
    This is the field that is integrated along the carriers' trajectories to calculate the induced signal. 
    }
    \label{fig:slices}
\end{figure*}

This paper presents the results of simulations for a single detector geometry as described by Table \ref{tab:params}. 
We use periodic boundary conditions in the $x$ and $y$ directions and a grounded plane beyond the anode at $z = 2d$, where $d$ is the detector thickness. 
The last boundary condition is imposed to simulate the potential inside and outside of the detector so that the electric field between pixels can be calculated.
The results for this simulation are shown in Fig. \ref{fig:slices}. 

\subsection{Charge Tracking and Integration} \label{tracking}
For each event, the {\tt Geant4} simulations give us the locations of the ionization energy deposits in the detector, which we offset to randomize the incident photon over a rectangular unit cell centered on a single pixel. 
We use this information to generate electrons and holes that are free to move through the detector. 
Energy depositions which occur sufficiently close to one another are merged and treated as a single deposition (discussed further in \ref{diff-rep}) while unmerged depositions are treated independently.

Subsequently, we track the motion of these electrons and holes through the detector accounting for the local electric field (caused by the applied detector bias), as well as charge carrier diffusion (a stochastic process) and electron-electron 
repulsion. 
We group charge carriers into small bundles (also referred to in this paper as ``charge elements'') which are tracked as a single unit, with electron and hole bundles being pulled in opposite directions by the local electric field, and individual charge bundles experiencing different diffusion and repulsion related displacements. 
The photo-electrons of some events cause the emission 
of fluorescent photons, that, when being absorbed, give rise to secondary photo-electrons. 
Our treatment allows us to track the free charge carriers from all these processes.   

\subsubsection{Carrier Drift}\label{drift}

We evolve the drift velocity with the equation
\begin{equation}
	\mathbf{v}_d = \pm\mu\mathbf{E} = \mp\mu\mathbf\nabla\phi,
    \label{eq:driftVelocity}
\end{equation}
where $\mu$ is the carrier mobility and is defined to be positive for both electrons and holes.
The upper signs are for holes and the lower signs are for electrons.
Table \ref{tab:params} lists the mobilities used in the simulations with subscripts $e$ and $h$ for electrons and holes, respectively. We calculate the electric field at the grid points as described above and
use a trilinear interpolation to calculate the electric field between grid points. The resulting change in position per integration step is calculated using 
Euler's method: $\delta\mathbf{r} = \mathbf{v}_d \delta t$.

\begin{table*}[tbh]
    \centering
    \begin{tabular}{|p{4.4cm}|p{2.4cm}|p{2.4cm}|p{4cm}|}
        \hline 
        Simulation parameter & Symbol & Value & Reference\\
        \hline
        Grid spacing in $x$, $y$ & $\delta x = \delta y$ & \SI{1/750}{mm} & \\
        Grid spacing in $z$ & $\delta z$ & \SI{1/50}{mm} & \\
        Dielectric constant &$\epsilon$ & 10 $\epsilon_0$& \cite{2011SPIE.8145E..07K}\\
        Time step & $\delta t$ & \SI{0.5}{ns} & \\
        Mean ionization energy & $\mathcal{E}$ & \SI{4.6}{eV} & \\
        Fano factor & $F$ & 0.089 &
        \cite{redus_pantazis_huber_jordanov_butler_apotovsky_1997} \\
        Electron mobility & $\mu_e$ & \SI{1000}{cm^2/V\ensuremath{\cdot}s} & \cite{nihCdTeconstants}\\
        Electron trapping time & $\tau_e$ & \SI{2.7}{\upmu s} & \cite{nihCdTeconstants}\\
        Hole mobility &$\mu_h$ & \SI{110}{cm^2/V\ensuremath{\cdot}s} & \cite{nihCdTeconstants}\\
        Hole trapping time & $\tau_h$ & \SI{1.6}{\upmu s}& \cite{nihCdTeconstants} \\
        Electron diffusion constant &$D_e$ & \SI{26}{cm^2/s} & \\
        Hole diffusion constant &$D_h$ & \SI{3}{cm^2/s} & \\
        Detector thickness & -- & \SI{1}{mm} & \\
        Pixel pitch & -- & \SI{150}{\upmu m} & \\
        Pixel spacing & -- & \SI{15}{\upmu m} & \\ 
        Cathode bias & -- & \SI{-150}{V} & \\ \hline
    \end{tabular}
    \caption{Constants and detector parameters for the simulation.}
    \label{tab:params}
\end{table*}

\subsubsection{Diffusion and Repulsion}\label{diff-rep}
We implement the expansion of a carrier cloud through diffusion and repulsion following Benoit and Hamel \citep{BENOIT2009508}. 
We neglect the attraction between charge carriers of opposite polarity, which may slow down the initial charge separation.
For an energy deposition $E_d$ and mean ionization energy per electron hole pair $\mathcal{E}$, an average of $N_0 = E_d /\mathcal{E}$ electron-hole pairs are created. The actual number of electron-hole pairs  created $N$ is chosen from a Gaussian distribution with mean $N_0$ and $\sigma = \sqrt{N_0 F}$, where F is the Fano factor. For CZT, this is $F = 0.089$ (Table \ref{tab:params}). To improve performance, the cloud of $N$ charge carriers is broken down into $n$ charge elements, each containing $N/n$ charge carriers. The initial distribution of these charge elements is modeled by a spherically symmetric Gaussian with RMS radius 
\begin{equation}
	R_p = A E_d \left(1 - \frac{B}{1 + C E_d} \right),
    \label{eq:Rp}
\end{equation}
where $A = \SI{0.95}{\upmu m/keV}$, $B = 0.98$, and $C = \SI{0.003}{keV^{-1}}$ are material-specific constants \citep{RevModPhys.56.S1}. 
In our simulations, this is not only used to parameterize the initial charge distribution, but it is also used to determine whether two clouds are close enough to affect each others' diffusion and repulsion, in which case the energy depositions are merged and the two clouds are approximated by a single cloud located at the center of mass.
The merging criteria was determined based on observations of the degree to which charge clouds expanded from the initial distribution in the same simulations. 

The density gradient within a cloud causes diffusion of the charge elements and the electric field created by the charges causes repulsion. These affect the charge density $\rho$ according to 
\begin{equation}
	\frac{\partial\rho}{\partial t} = D\nabla^2\rho - \mu\nabla\cdot(\rho\mathbf{E}),
    \label{eq:diffusion_repulsion}
\end{equation}
where $D$ is the diffusion coefficient, $\mu$ is the carrier mobility, and $\mathbf{E}$ here is the electric field produced by the charges in the cloud \citep{BENOIT2009508}. 
The values of $D$ used in our simulation for electrons and holes are listed in \ref{tab:params} with subscripts $e$ and $h$, respectively, and were derived using the Einstein relation for charged particles at room temperature:
\begin{equation}
    D = \frac{\mu k_B T}{q}.
\end{equation}
Benoit and Hamel showed that for an ellipsoidal Gaussian charge distribution with $\boldsymbol\sigma = (\sigma_x, \sigma_y, \sigma_z)$, the time evolution of the distribution is described by 
\begin{equation}
	\frac{\partial\boldsymbol\sigma(t)^2}{\partial t} = 2\mathbf{D}',
    \label{eq:sigSq}
\end{equation}
where $\mathbf{D}'$ is the vectorial effective diffusion coefficient
\begin{equation}
	\mathbf{D}' = \left(D + \lambda \frac{\sigma_x}{\sigma_y\sigma_z}, D + \lambda \frac{\sigma_y}{\sigma_x\sigma_z}, D + \lambda \frac{\sigma_z}{\sigma_x\sigma_y}\right),
    \label{eq:effectiveDiffusion}
\end{equation}
with:
\[
\lambda \equiv \frac{eN\mu}{20\sqrt5\pi\epsilon}.
\]
This effective diffusion coefficient encapsulates the effects of both diffusion and repulsion. 
At each time step, $\boldsymbol\sigma$ and $\mathbf{D}'$ are recalculated and each charge element is displaced by a random walk in $x$, $y$, and $z$ by distances selected from a Gaussian parameterized by $\sigma_i = \sqrt{2D'_i\delta t}$, where $\delta t$ is the size of the time step.

\subsubsection{Trapping}
While the charge carriers are moving in the detector, they may encounter impurities and recombine in the material, or are trapped and released on time scales longer than the shaping time of the readout electronics. When this occurs, we effectively lose the charge as it no longer contributes to the signal. We account for trapping by reducing the amount of charge present according to:
\begin{equation}
    q(t) = q_0 e^{-t/\tau},
\end{equation}
where $q_0$ is the initial charge at time $t=0$, $t$ is the time since the energy was deposited, and $\tau$ is the carrier lifetime, 
which quantifies trapping as a bulk property of the cloud. The values of $\tau$ used in the simulations are listed in Table \ref{tab:params} with subscripts $e$ and $h$ electrons and holes, respectively.
Trapping is caused by crystal defects such as Te precipitates
\citep[e.g.][]{2017SPIE10423E..1MB,2018SPIE10762E..0VZ,2019NIMPA.924...28W,2019NIMPA.923...51K,2019SPIE11114E..1NY,2020NIMPA.95861996B,2020JCrGr.53525542M}. Owing to the lack of information about the spatial distribution and size of charge trapping sites, we neglect the statistical fluctuations caused by the 
charge trapping mechanisms.
Detailed comparisons of simulations as those presented here and measured data could be used to constrain these material properties.
\subsubsection{Charge integration} \label{integration}

Charged particles moving near an electrode induce a charge on the electrode. Shockley showed a way to quantify this with the Shockley-Ramo Theorem \citep{shockley,zhong}:
\begin{equation}
    q_{ind} = -q \Delta \phi_{wp},
    \label{eq:shockley_charge}
\end{equation}
where $\Delta\phi_{wp}$ is the change in weighting potential between two points in the charge $q$'s trajectory and $q_{ind}$ is the charge induced on the electrode.
The weighting potential is a unit-less potential found by setting the electrode for which we want to calculate the induced charge to 1 and all others to 0. 
With these boundary conditions, the procedure described in \ref{pot} is followed to calculate the weighting potential.
The right panels of Figure \ref{fig:slices} show that most of the change in weighting potential occurs near the pixels. 
This is the well-known small pixel effect \citep{PhysRevLett.75.156, zhong}:
as the signal is mostly generated near the pixels, it becomes largely independent of the location of the primary interaction.

To simulate the currents read by the electrodes, we use an alternate form of Eq. \eqref{eq:shockley_charge}:
\begin{equation}
    I_{ind} = q \mathbf{v} \cdot \mathbf{E}_{wp},
    \label{eq:shockley_current}
\end{equation}
where $\mathbf{v}$ is the instantaneous velocity of the charge (considering both drift and diffusion) computed from the change of location
in this time step, and $\mathbf{E}_{wp}$ is the ``electric field'' resulting from the weighting potential. These are implemented as the average velocity over the whole time step and the trilinear interpolated weighting potential.

For each event, there may be several energy depositions following the initial photon interaction in the crystal, 
including elastic and inelastic photon scattering, and the ionization along the track of energetic electrons.
Following the design of the HEXID ASIC, we assume that the electronics read out all pixels with an energy exceeding the trigger threshold as well as the immediate neighbors of these pixels (even if the signals in the neighboring pixels do not exceed the trigger threshold). We will see below that this scheme leads to excellent energy resolutions at lower photon energies ($<\sim$\SI{60}{keV}). At higher energies, energy escaping the detector or traveling to pixels further away starts to become an issue, and does deteriorate the detectors' energy resolutions.
We analyzed the data for 
% both the cases of the unipolar integration (readout of the negative charge signal only), and 
the bipolar integration of the signal (read out of the integrated negative and positive charge).

%%%%%%%%%%%%%%%%%%%%%%%%%%%%%%%%%%%%%%%%%%%%%%%%

\section{Results} \label{results}

%%%%%%%%%%%%%%%%%%%%%%%%%%%%%%%%%%%%%%%%%%%%%%%%

In this section, we discuss methods to reconstruct the energy of the incident photon, and the location of the primary interaction. We will first discuss the results obtained in the absence of  readout noise, and the show how they  change as we add the noise expected for
the HEXID ASIC.

\subsection{Energy reconstruction}

The procedure in Section \ref{method} was followed for a detector with the boundary conditions described in \ref{pot} and with \SI{22.9}{keV} incident photons.

\begin{figure}
    \centering
    \includegraphics[width=0.97\linewidth]{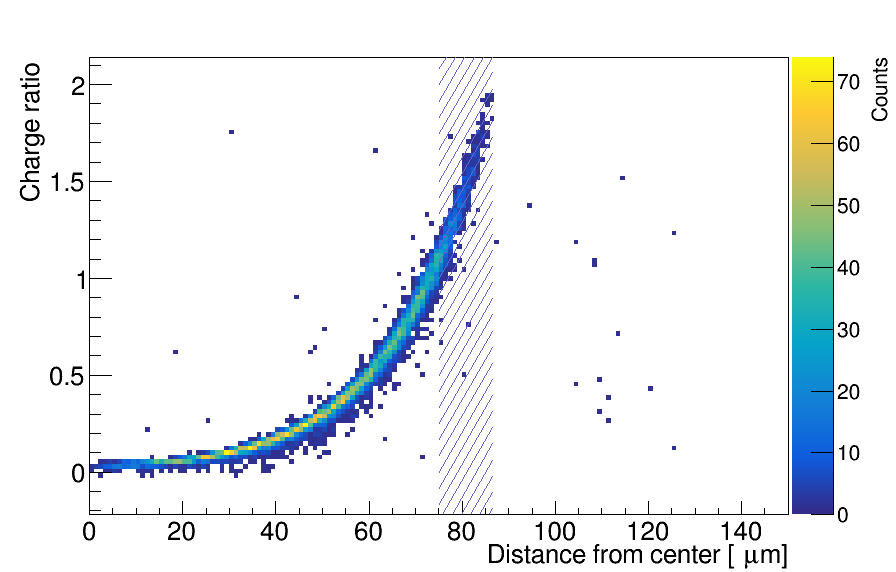}
    \caption{The panel shows the ratio $r$ of the charge induced on neighbor pixels divided by the charge induced on the brightest pixel
    (Eq. \eqref{eq:chargeRatio}) as a function of the distance of the initial charge deposition from the center of the brightest pixel
    for incident 22.9 keV photons. 
    The initial photon positions are randomized over a rectangular unit cell. 
    Due to the hexagonal pixel geometry, the pixel edge may be located between $1/2$ and $1/\sqrt3$ pixel pitches from the center, indicated by the shaded area. 
    As expected, there is more charge sharing as the interaction occurs farther from the pixel center.
    }
    \label{fig:distToRatio}
\end{figure}

We characterize the amount of charge sharing with the charge ratio:
\begin{equation}
    r = \frac{Q_{neighbors}}{Q_{center}},
    \label{eq:chargeRatio}
\end{equation}
where $Q_{center}$ is the charge induced on the center pixel (the pixel with the largest signal), and $Q_{neighbors}$ is the total charge induced on its six neighbors. As expected, the charge ratio $r$ increases with
the distance from the center of the
pixel (Fig. \ref{fig:distToRatio}).

\begin{figure}
    \centering
    \includegraphics[width=0.97\linewidth]{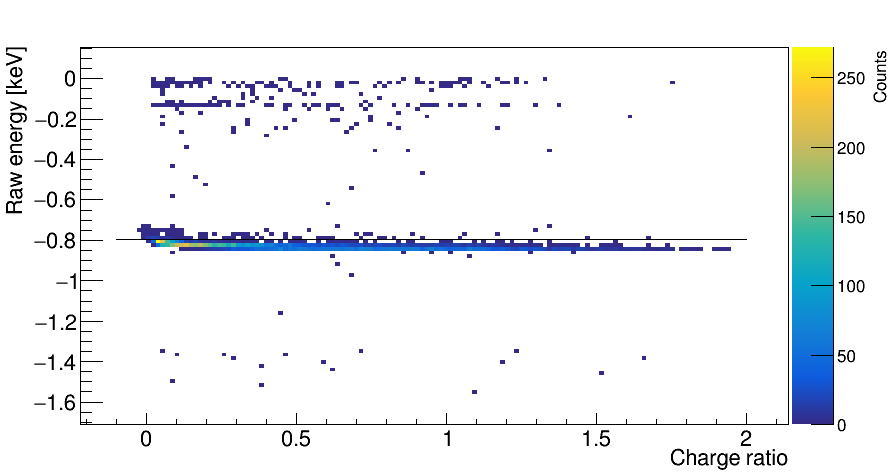}
    \caption{The total reconstructed charge for each event as a function of the charge ratio. 
    The horizontal line indicates the expected total charge from a \SI{22.9}{keV} incident photon.
    On average, there is more induced charge than expected. 
    This small discrepancy may be due to the next layer of neighbors which were not considered; they are far enough that an opposite sign charge could have been induced, leading to more charge appearing to be induced within the nearest neighbors.}
    \label{fig:ratioCharge}
\end{figure}

We see that adding up the charge from the center pixel which exceeds a certain threshold value and all its adjacent pixels 
is important for achieving good energy resolution. This requirement means that
the total readout noise is given by the quadratic sum of the readout noise of a total of between one and seven pixels. 
Fig. \ref{fig:ratioCharge} shows the total induced charge on the center pixel and its nearest neighbors plotted against the charge ratio. 
The sum signal decreases (increases in magnitude) with increasing charge ratio, which may be indicative of small negative induced charges from farther pixels for which the signal was not calculated.

To correct for this difference in charge, a multiplicative correction histogram was made from applying cuts at $-0.6$ and \SI{-1}{fC} to the plot in Fig. \ref{fig:ratioCharge} and taking the average of the remaining points for each charge ratio bin. 
For each event, the reconstructed energy is corrected by multiplying the raw integrated charge by the value of this correction histogram corresponding to the event's charge ratio.

\begin{figure}
    \centering
    \includegraphics[width=0.95\linewidth]{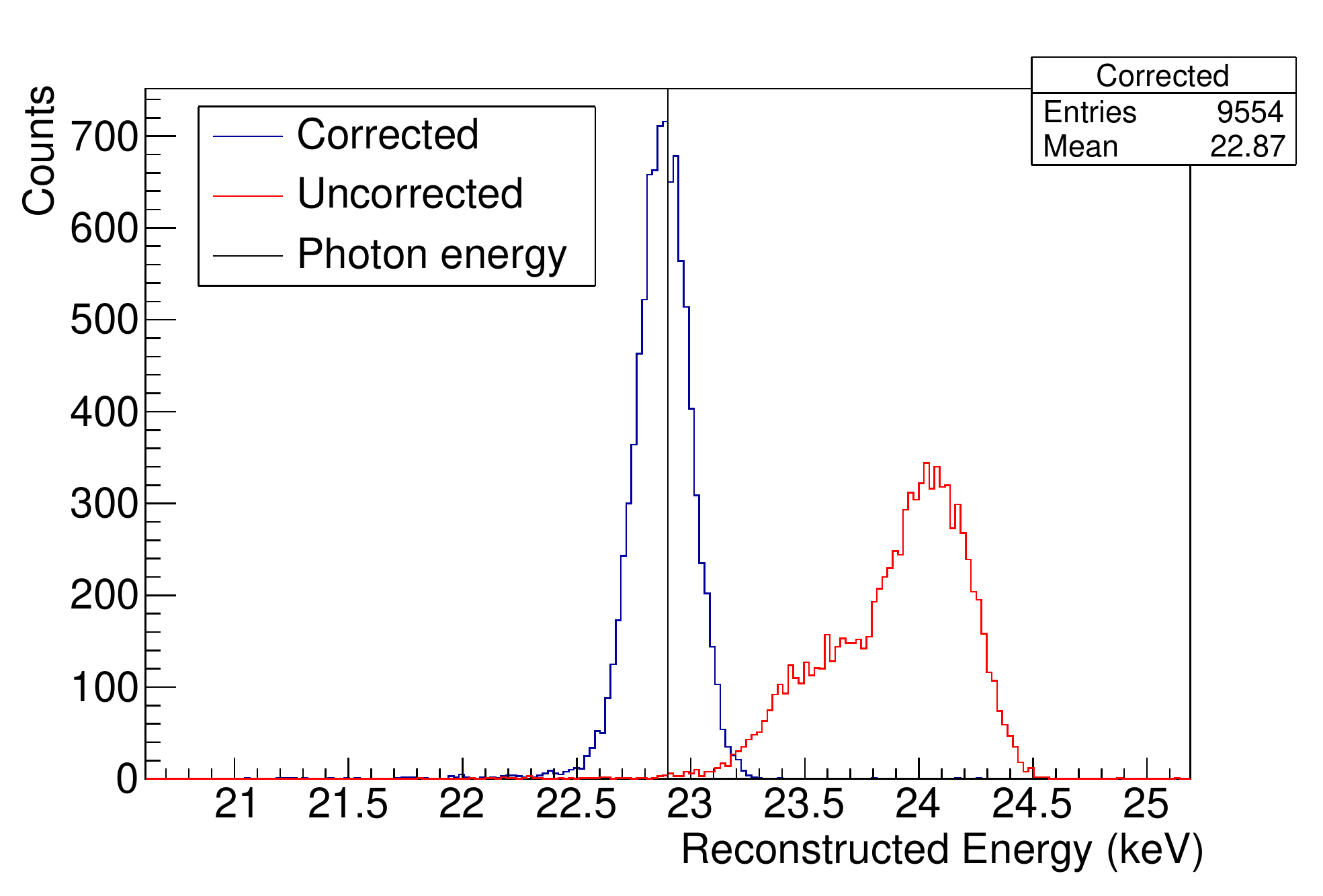}
    \caption{The reconstructed energies from the simulations for \SI{22.9}{keV} incident photons. The red histogram shows the raw reconstructed energies inferred from summing the signal in the brightest pixel and its neighbors. The blue histogram shows the energies after correcting the signal with the charge ration $r$ from Equation \eqref{eq:chargeRatio}. The black line indicates the incident photon energy.}
    \label{fig:energyReconstruction}
\end{figure}

Figure \ref{fig:energyReconstruction} shows the energies reconstructed from charge integration both before and after multiplicative correction. The corrected peak has a full width at half maximum (FWHM) of \SI{0.260}{keV}. The energy spectrum shown here does not yet include the electronic readout noise.

The readout Equivalent Noise Charge (ENC) depends on the detector capacitance and the readout electronics. The pixel capacitance can be calculated from two
maps of the potential calculated for different bias voltages.
Integrating the electric field around the pixel gives the charge on the pixel via 
Gauss' law. The difference $\Delta V$ of the two bias voltages  and the
difference $\Delta q$ of the inferred charges gives the capacitance 
according to:
\begin{equation}
    C = \frac{\Delta q}{\Delta V}.
\end{equation}

\begin{figure}
    \centering
    \includegraphics[width=0.95\linewidth]{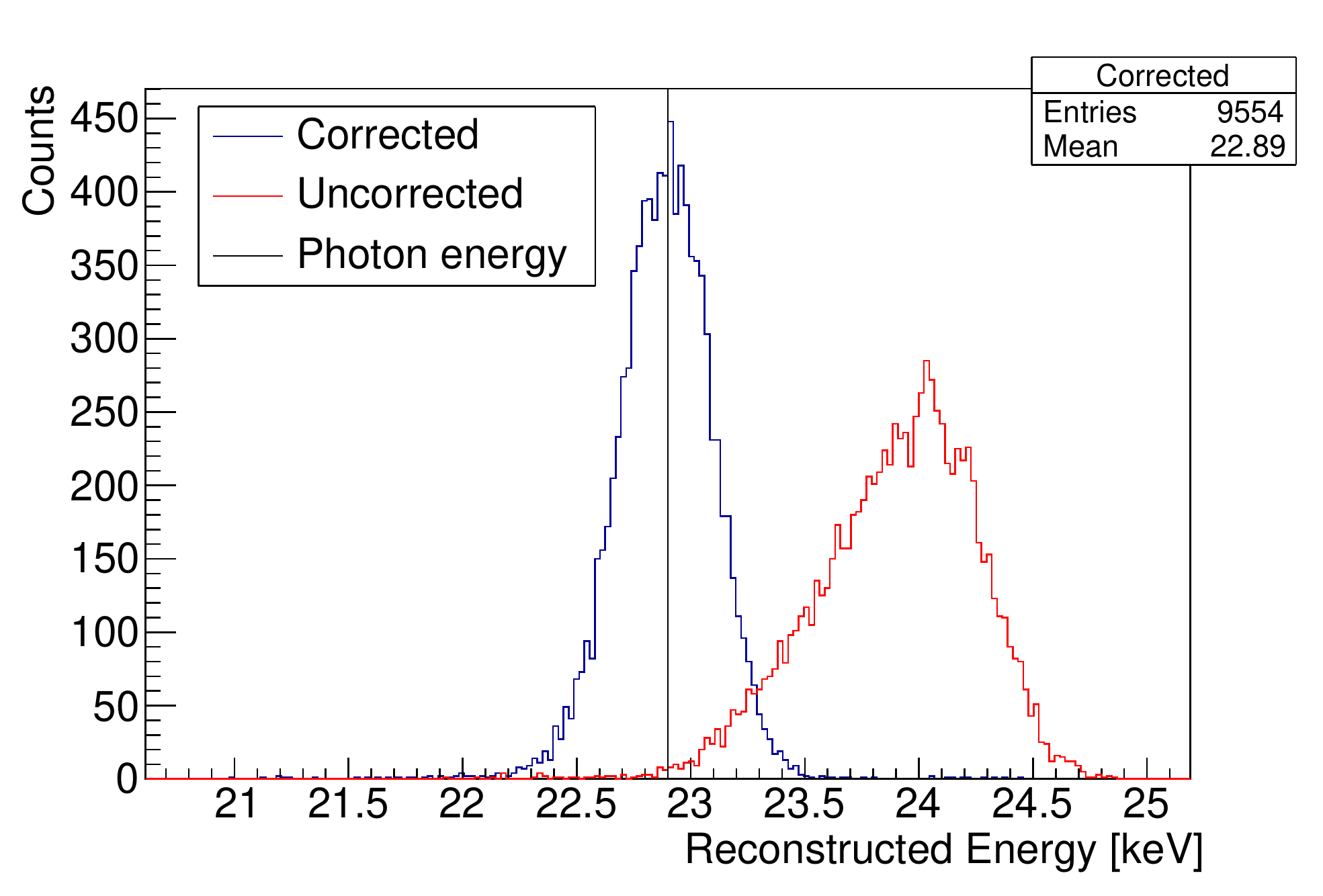}
    \caption{The reconstructed energies with a 14 electron ENC. The same correction histogram as that used for Fig. \ref{fig:energyReconstruction} was used.}
    \label{fig:energyNoise_22}
\end{figure}

For this \SI{1}{mm} thick CZT detector with with \SI{150}{\upmu m} pixels, the per-pixel capacitance is \SI{6.94}{pF}. 
For the HEXID architecture, this capacitance leads to a conservatively-estimated noise of 14-electron ENC. 
Figure \ref{fig:energyNoise_22} shows the results of the energy reconstruction with this noise included. 
The FWHM is \SI{0.435}{keV}. 
The same correction histogram as before was used. 
The corrected energies' peak is still centered at the photon energy, which is expected given that the Gaussian was centered about 0,  and that the noise is small compared to the total charge across all pixels, which contributes to minimal shifts in the charge ratio and therefore the correction factor.

\begin{figure}
    \centering

    \includegraphics[width=0.95\linewidth]{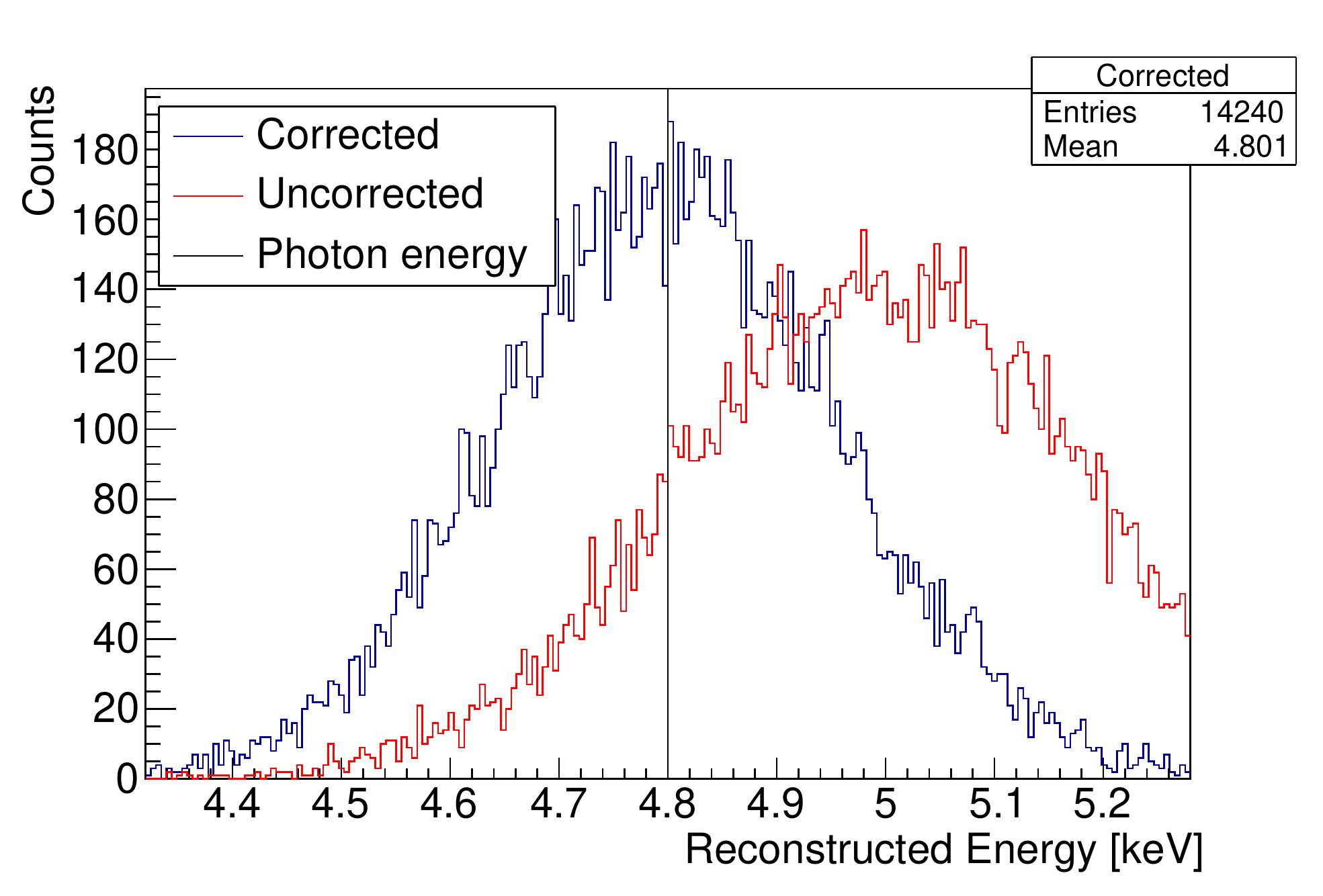}
    
    \includegraphics[width=0.95\linewidth]{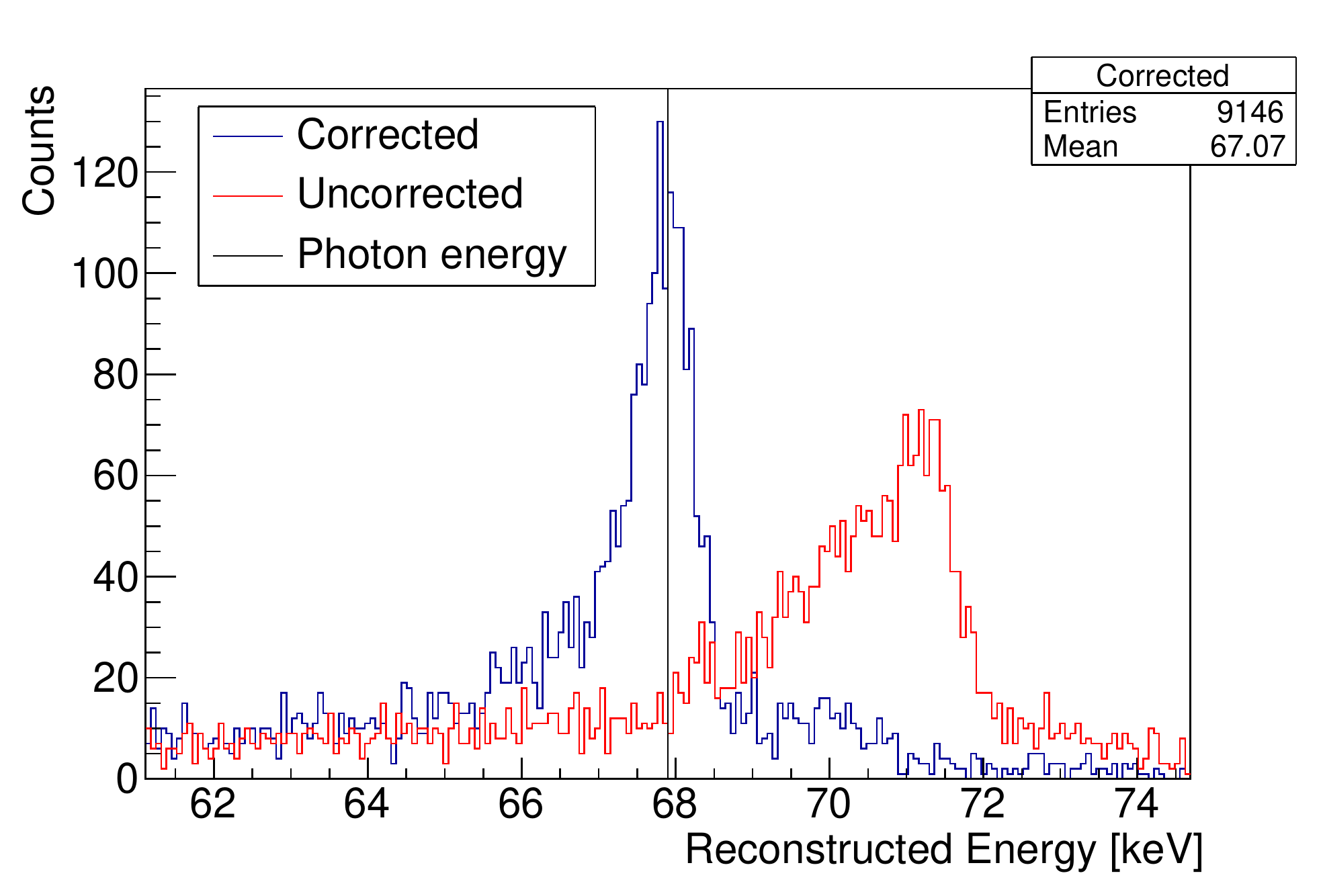}
    
    \includegraphics[width=0.95\linewidth]{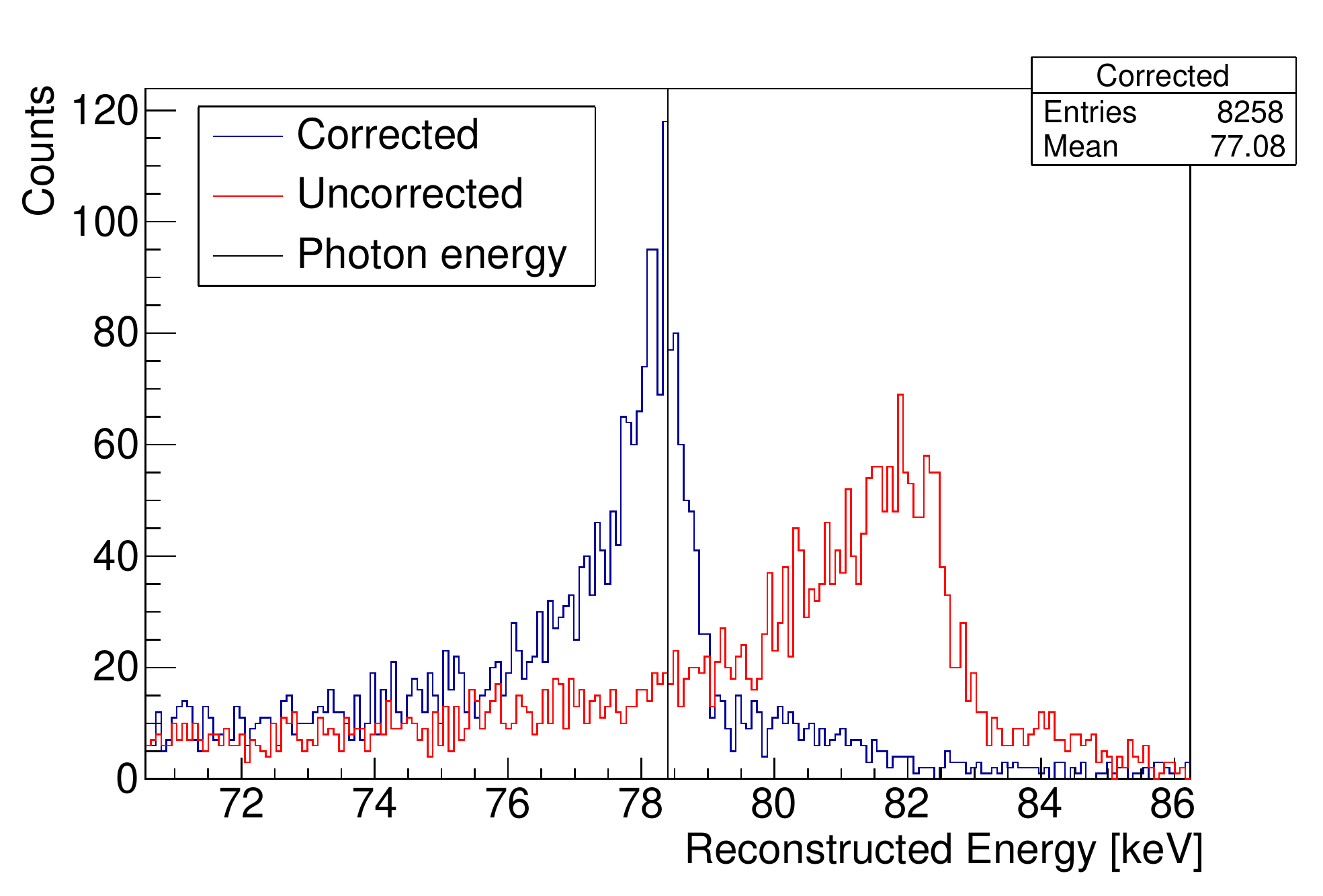}
    
    \caption{The reconstructed energies for \SI{4.8}{keV} (\emph{top}), \SI{67.9}{keV} (\emph{middle}), and \SI{78.4}{keV} (\emph{bottom}) photon events with 14-electron ENC. All used the correction histogram from the \SI{22.9}{keV} simulations to correct the reconstructed energy based on the charge ratio and include a 14-electron ENC on each pixel. The \SI{67.9}{keV} and \SI{78.4}{keV} events were multiplied by an additional correction factor to center the peak on the incident photon energy without significantly impacting the energy resolution.}
    \label{fig:otherEnergies}
\end{figure}

We tested if a single correction histogram (derived for \SI{22.9}{keV} photons) can be used to reconstruct the initial photon energy for
a wide range of initial photon energies, i.e. \SI{4.8}{keV}, \SI{67.9}{keV}, and \SI{78.4}{keV}. 
Figure \ref{fig:otherEnergies} shows that a single correction histogram indeed gives excellent results for all these photon energies.
For \SI{4.8}{keV} photon events with noise, the corrected energy peak has a FWHM of \SI{0.350}{keV}. 
For \SI{67.9}{keV} and \SI{78.4}{keV} photons, the corrected peak was slightly below the incident photon energy.
The reconstructed energies were multiplied by an additional correction factor to push the peak of the energy spectrum to the true X-ray energy. 
These factors were less than 1\%, so they did not significantly impact the energy resolution.
Note that the average values shown in the figures are 
lower than peak location due to the presence of the 
low-energy tails.
 The FWHM calculated from the binned spectra were 
\SI{0.815}{keV} for the \SI{67.9}{keV} incident photons and \SI{1.333}{keV} for the \SI{78.4}{keV} photons.
The energy resolutions due to individual sources of noise are summarized in Table \ref{tab:widths}.

\begin{table}
    \centering
    \begin{tabular}{|p{1.2cm}|p{1.2cm}| p{1.2cm}| p{1.2cm}| p{1.4cm}|}
         \hline 
         Photon energy [keV] &  Fano [keV]& Location Dependence [keV]  & ENC [keV]& Total [keV]\\
         \hline
         4.8 & 0.104 & 0.017 & 0.334 & 0.350 \\
         22.9 & 0.228 & 0.125 & 0.349 & 0.435 \\ 
         67.9 & 0.398 & 0.376 & 0.607 & 0.815 \\
         78.4 & 0.421 & 1.014 & 0.756 & 1.333 \\ 
         122.1 & 0.525 & 3.995 & 2.421 & 4.701 \\
         158.4 & 0.598 & 7.818 & 0 & 7.841 \\
         \hline
    \end{tabular}
    \caption{The resolution (FWHM) of the reconstructed energy from each source of noise: Fano noise, dependence of the induced charge on the locations of the energy deposition, electrical readout noise (ENC), and total noise. 
    The \SI{158.4}{keV} resolutions are calculated using twice the right-sided half width at half maximum due to high counts at low energies outside the photopeak.
    The contribution of ENC to the \SI{158.4}{keV} peak resolution was not significant at the binning for which the peak was resolvable. }
    \label{tab:widths}
\end{table}

\begin{figure}
    \centering
    \includegraphics[width=0.95\linewidth]{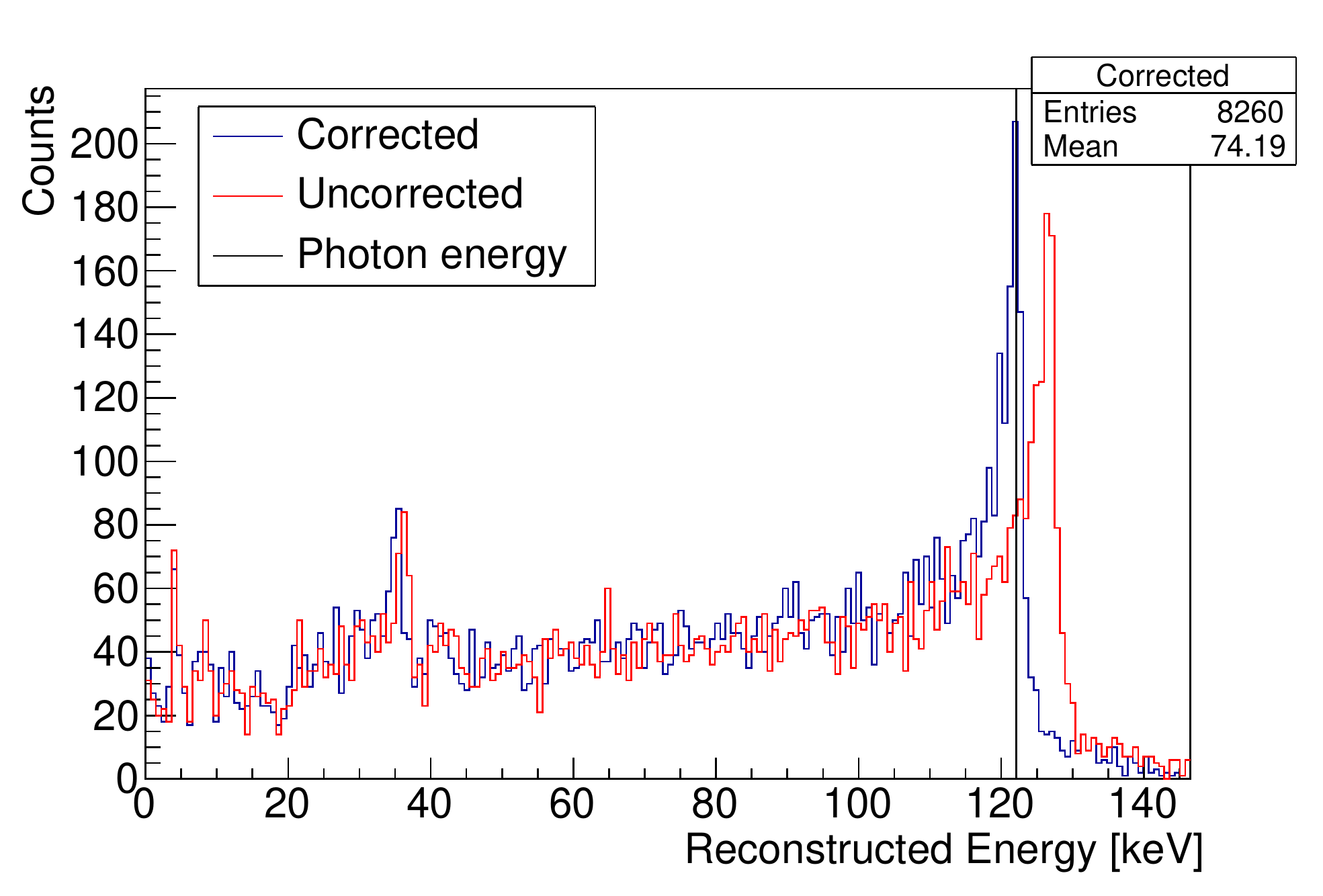}
    
    \includegraphics[width=0.95\linewidth]{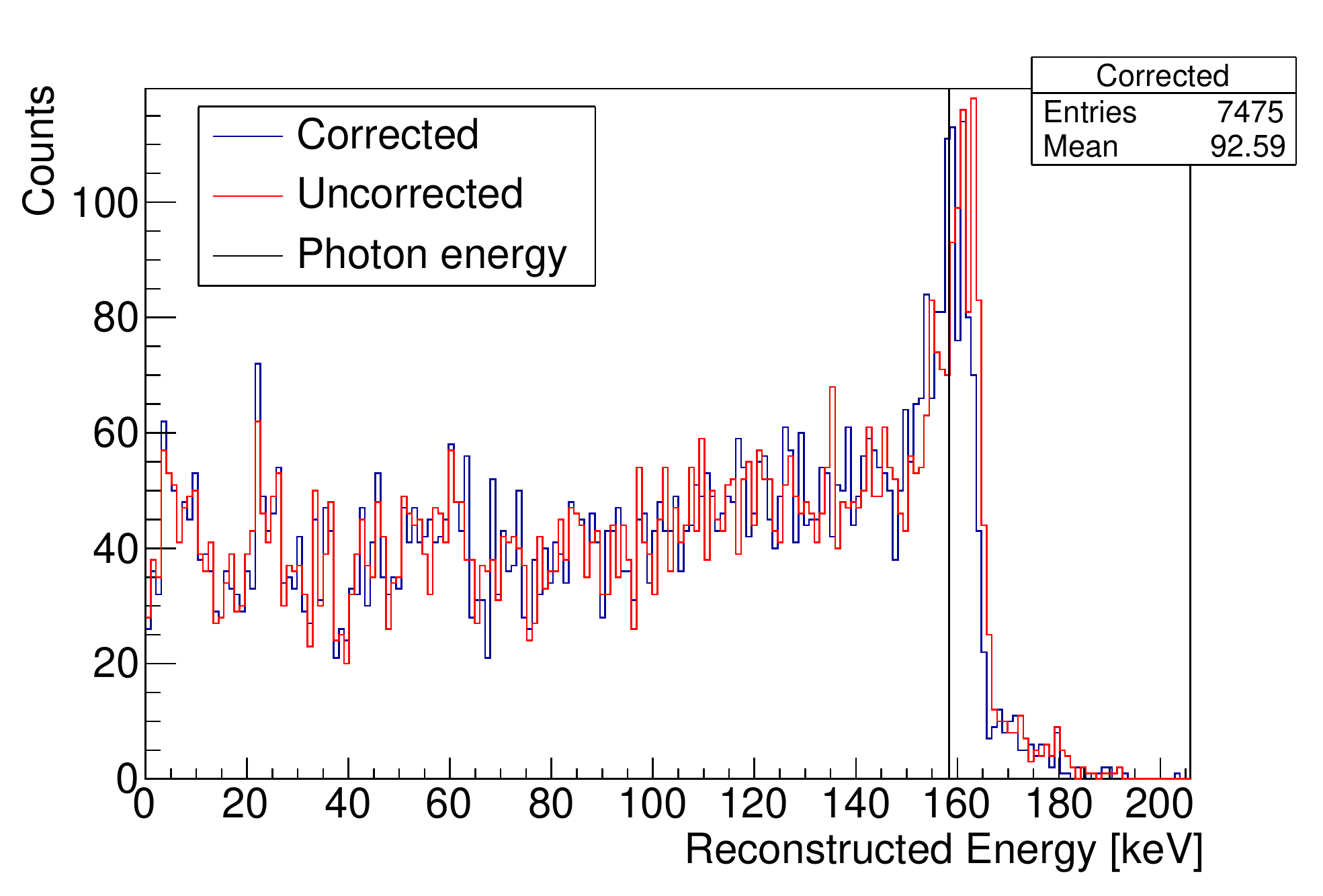}
    \caption{The reconstructed energies for \SI{122.1}{keV} (\emph{top}) and \SI{158.4}{keV} (\emph{bottom}) photon events with 14-electron ENC. There are significant energy losses as evidenced by the high level of reconstructed energies down to \SI{0}{keV}. These reconstructed energies were multiplied by a small ($\sim$1) correction factor to shift the peak to the true photon energy.  }
    \label{fig:veryHighEnergy}
\end{figure}

\begin{figure*}
    \centering
    \textbf{(a)}
    \includegraphics[width=0.6\linewidth/2, valign=t]{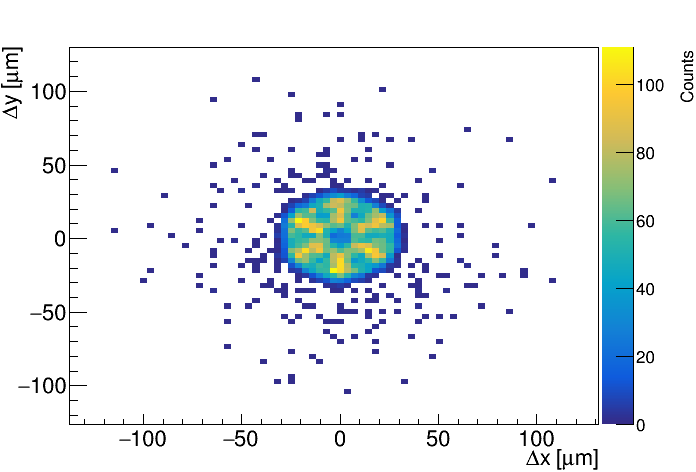}
    \hspace{0.5cm}
    \includegraphics[width=0.6\linewidth/2, valign=t]{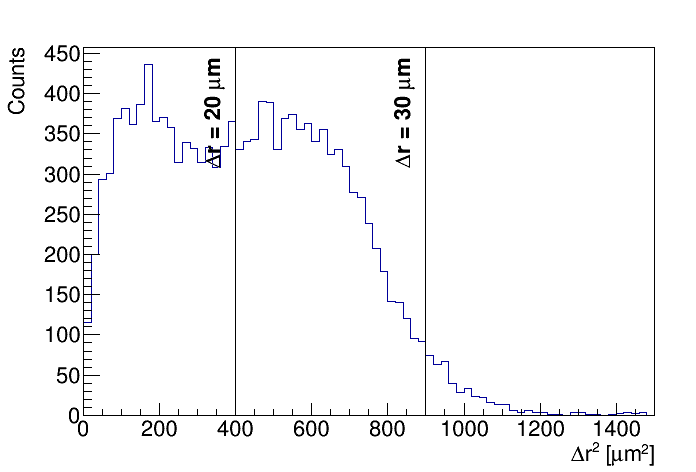}
    
    \textbf{(b)}
    \includegraphics[width=0.6\linewidth/2, valign=t]{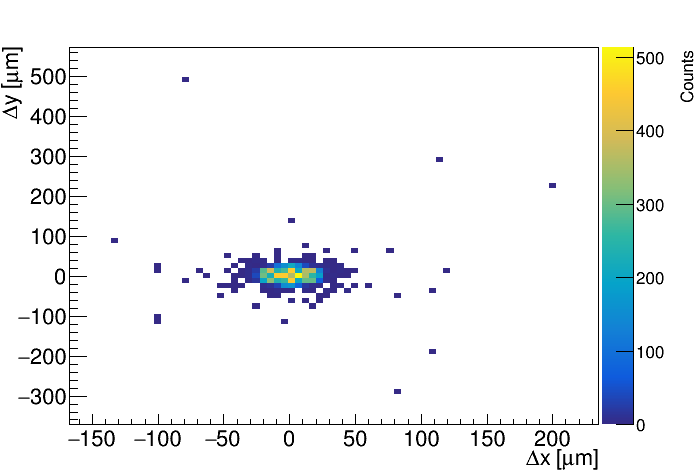}
    \hspace{0.5cm}
    \includegraphics[width=0.6\linewidth/2, valign=t]{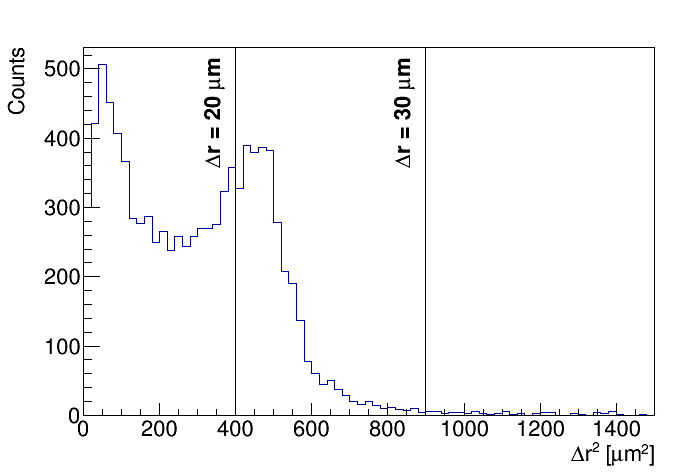}
    
    \textbf{(c)}
    \includegraphics[width=0.6\linewidth/2, valign=t]{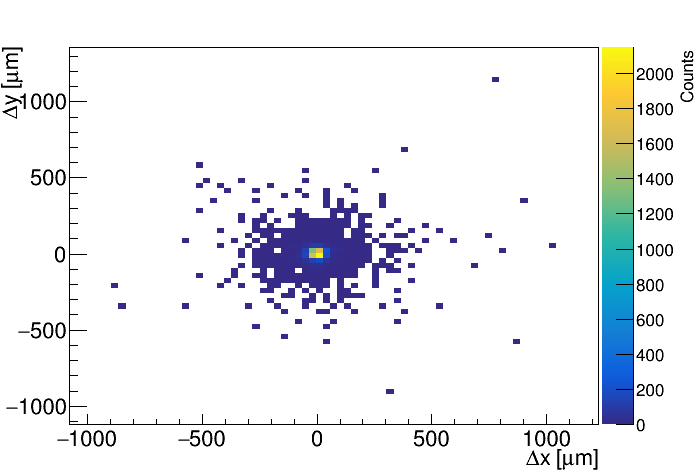}
    \hspace{0.5cm}
    \includegraphics[width=0.6\linewidth/2, valign=t]{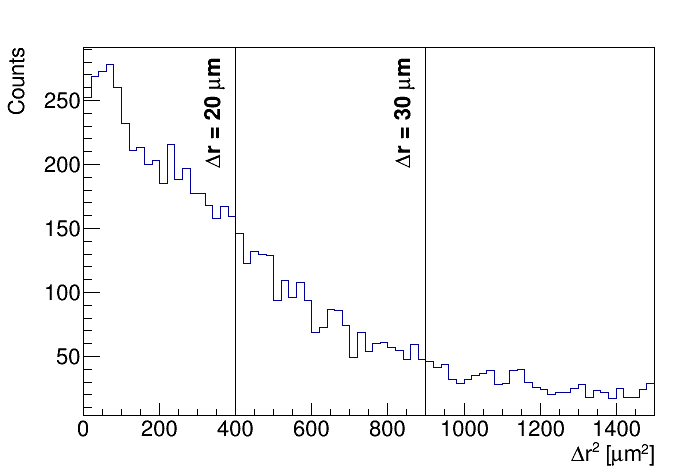}
    
    \textbf{(d)}
    \includegraphics[width=0.6\linewidth/2, valign=t]{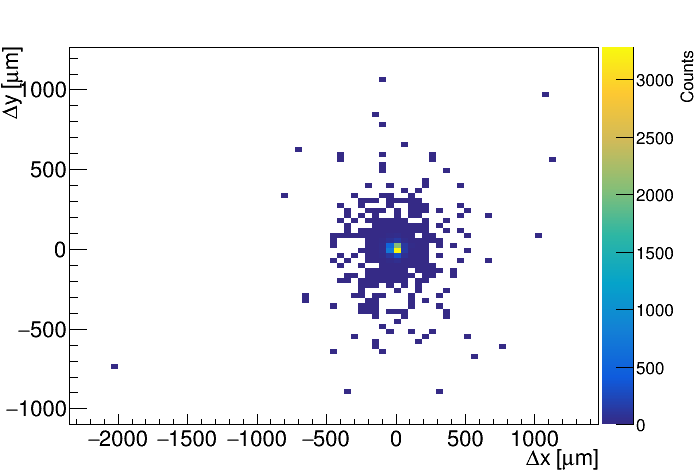}
    \hspace{0.5cm}
    \includegraphics[width=0.6\linewidth/2, valign=t]{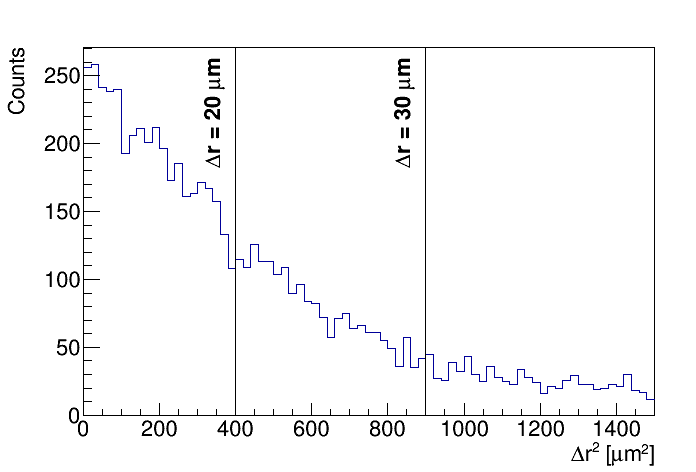}
    
    \textbf{(e)}
    \includegraphics[width=0.6\linewidth/2, valign=t]{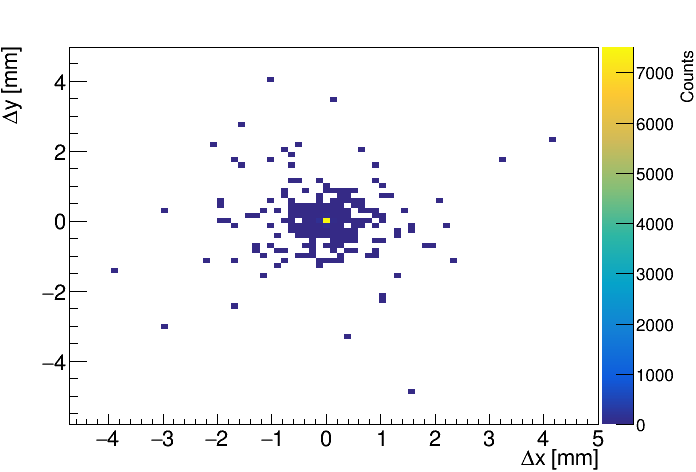}
    \hspace{0.5cm}
    \includegraphics[width=0.6\linewidth/2, valign=t]{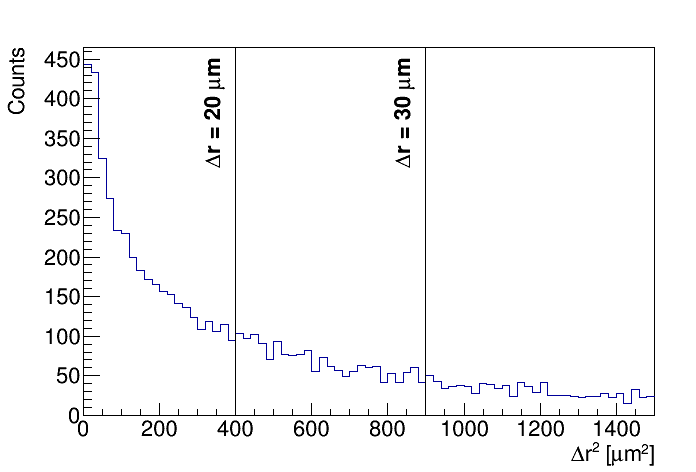}
    
    \textbf{(f)}
    \includegraphics[width=0.6\linewidth/2, valign=t]{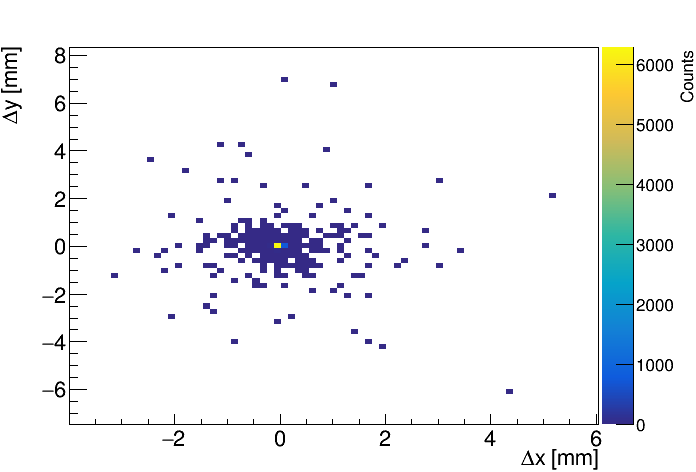}
    \hspace{0.5cm}
    \includegraphics[width=0.6\linewidth/2, valign=t]{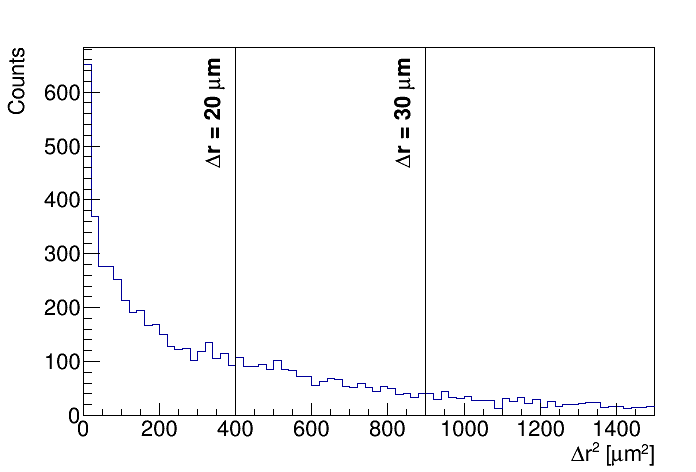}
    
    \caption{The absolute error in the reconstructed $x$ and $y$ position of the incident photon using the energy-weighted average of the integrated charge on the brightest pixel and its nearest neighbors (\emph{left}) and the squared distance $\Delta r^2$ between the reconstructed position and incident photon (\emph{right}) for \SI{4.8}{keV} (a), \SI{22.9}{keV} (b), \SI{67.9}{keV} (c), \SI{78.4}{keV} (d), \SI{122.1}{keV} (e), and \SI{158.4}{keV} (f) photons. 
    Some reconstructed positions which are farther away from the main cluster may be due to effects from noise, as in the \SI{4.8}{keV} case, and scattering in the higher energies.  
    Lines are drawn at the locations of $\Delta r = 20$ and \SI{30}{\upmu m}. }
    \label{fig:reconstructedPositions}
\end{figure*}

At even higher energies, the detector's performance starts to deteriorate markedly. 
Figure \ref{fig:veryHighEnergy} shows simulated results for the
\SI{122.1}{keV} and \SI{158.4}{keV} lines from $^{57}$Co and $^{56}$Ni, respectively.
The energy spectra show pronounced low-energy tails.
Two effects contribute to this effect: some of the photons 
interact close to the anode side of the detector, so that the drifting electrons induce little
charge before impinging on the anodes (see Fig. \ref{fig:slices}, Eq. \eqref{eq:shockley_charge}). 
The second effect is that scattered photons 
are only absorbed after traveling beyond the next neighbor pixels, or leave the detector altogether. 
In both cases, their energy does not count 
towards the reconstructed  energy.
The \SI{122.1}{keV} and \SI{158.4}{keV} energy resolutions (including all effects) are
\SI{4.7}{keV} and \SI{7.8}{keV} respectively (Table \ref{tab:widths}).
At these high energies, the correction with the charge ratio $r$ 
does not lead to a marked improvement of the energy resolutions.
We therefore only used the raw integrated charge and a peak-shifting multiplicative factor close to one to 
reconstruct these energies. Developing and testing a more sophisticated Maximum Likelihood 
energy reconstruction is outside of the scope of this paper.
Most photons do not interact with the detector at all; only 41\% of \SI{122.1}{keV} and 25\% of \SI{158.4}{keV} photons deposited any energy into the detector.
Of these, 15\% of \SI{122.1}{keV} and 14\% of \SI{158.4}{keV} events resulted in reconstructed energies in the photopeak.

%We have demonstrated a possible dependency of the charge measurement on the amount of charge sharing even for a single incident photon energy.  Our proposed use of a charge ratio-dependent correction histogram is blind to the the initial photon energy (though it effectively slightly over-corrects at higher energies) and accounts for the degree of charge sharing. A single correction histogram is also advantageous at higher incident photon energies due to the loss of charge, adding low-energy noise in the equivalent of Fig. \ref{fig:ratioCharge} which would result in a poorly-performing energy-dependent correction histogram or correction factor.

\subsection{Position reconstruction} \label{position_reonstruction}
In the next step, we reconstruct the original photon position using the charge-weighted average position of the brightest pixel and its six nearest neighbors. The simulations include the readout noise of 14-electron ENC.

\begin{figure}
    \centering
    \includegraphics[width=0.95\linewidth]{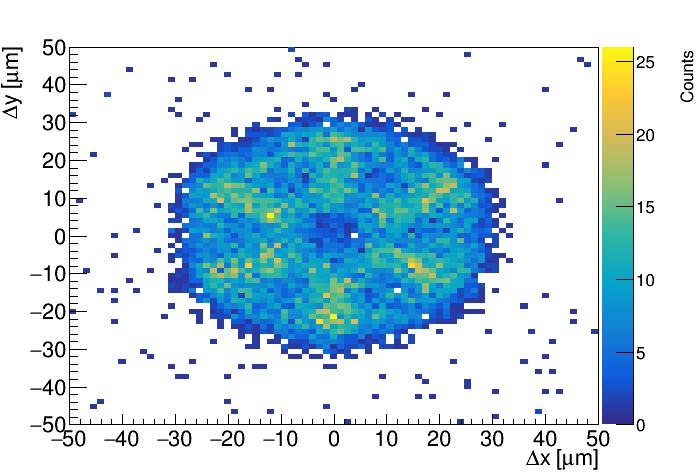}
    
    \includegraphics[width=0.95\linewidth]{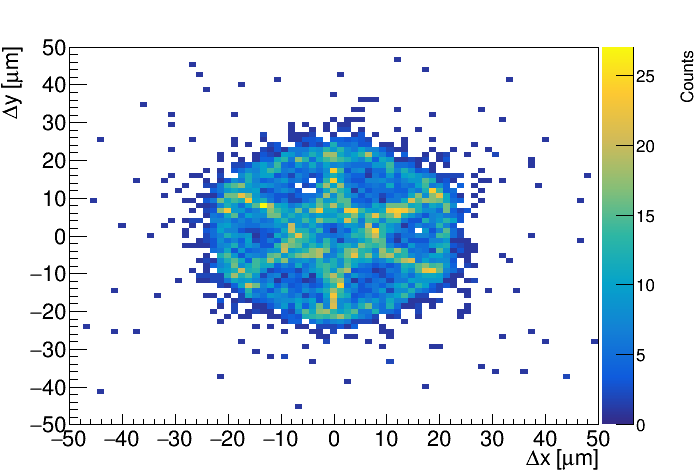}

    \caption{The errors in reconstructed photon positions 
    for \SI{4.8}{keV} (\emph{top}) and \SI{22.9}{keV} (\emph{bottom}) zoomed to \SI{\pm50}{\upmu m} in either direction. These account for over 99\% of all reconstructed positions at these energies.}
    \label{fig:positions_zoomed}
\end{figure}

Figure \ref{fig:reconstructedPositions} shows the absolute difference in the reconstructed $x$ and $y$ position of the incident photon using the weighted average of the pixels' center coordinates.
We first discuss the results for \SI{4.8}{keV} and \SI{22.9}{keV} photons.
There are clusters near $(0,0)$ for both energies that indicate that the reconstruction is generally accurate at these energies.
In the reconstructions for both energies, there are a few reconstructed positions that are considerably farther away ($>$1 pixel pitch) from the actual photon position. 
We used simulations without ENC to prove that the main cause of this effect is the ENC for the \SI{4.8}{keV} events and scattering for the \SI{22.9}{keV} events. 
Note that the number of outliers is small. 
For the \SI{22.9}{keV} events, 99.6\% of the reconstructed positions are within \SI{50}{\upmu m} from the coordinates of the incident photon in both the $x$ and $y$ directions;
for the \SI{4.9}{keV} events, this number is 99.1\%. Figure \ref{fig:positions_zoomed} zooms into this inner region and shows some interesting sub-structure. 

The situation is somewhat different at higher energies (Fig.\ \ref{fig:reconstructedPositions}, lower four panels).
The increased effect of charge sharing leads to a better position reconstruction for some events and thus to a more centrally peaked distance distribution. 
However, as scattered photons can travel farther away from the location of the first interaction, there is also a larger tail towards larger displacements. 
The net effect is that the HPD of the reconstructed positions is largely 
independent of the energy of the primary photon (Table \ref{tab:hpds}).

\begin{table}
    \centering
    \begin{tabular}{| >{\centering\arraybackslash}m{3cm} | >{\centering\arraybackslash}m{3 cm} |}
         \hline 
         Photon energy [keV] & Position Reconstruction HPD [$\upmu$m]\\
         \hline
         4.8 & 41.8 \\
         22.9 & 34.5 \\ 
         67.9 & 43.4 \\
         78.4 & 41.0 \\ 
         122.1 & 46.4 \\
         158.4 & 37.7 \\
         \hline
    \end{tabular}
    \caption{The spatial resolution (HPD) of the reconstructed positions from the simulated incident photons. }
    \label{tab:hpds}
\end{table}

%%%%%%%%%%%%%%%%%%%%%%%%%%%%%%%%%%%%%%%%%%%%%%%

\section{Discussion}\label{disc}
In this paper, we present simulations of \SI{1}{mm} thick CZT detectors with hexagonal pixels 
at an extremely small pixel pitch of \SI{150}{\upmu m}.  The detector simulations account
for the spatially distributed generation of free charge carriers in the detector, and the
drift and diffusion of the charge of both polarities. The simulations furthermore account
for the anticipated charge resolution of the HEXID3 ASIC.  We have shown that the sum of 
the signals of the brightest pixel and the adjacent pixels and the ratio of the sum of the charge in the neighbor pixel divided by the charge of the brightest pixel, 
give FWHM energy resolutions of 
\SI{350}{eV} at \SI{4.8}{keV},  
\SI{435}{eV} at \SI{22.9}{keV}
\SI{815}{eV} at \SI{67.9}{keV}, 
\SI{1.333}{keV} at \SI{78.4}{keV},
\SI{3.4}{keV} at \SI{122.1}{keV}, and
\SI{9.8}{keV} at \SI{158.4}{keV}.
The charge induced on the brightest pixels and its neighbors can be used to reconstruct the location of the energy deposition to an accuracy of better than 50$\mu$m for all
the energies we simulated (4.8-\SI{158.4}{keV}).
The simulations were performed using only the brightest pixel and its nearest neighbors to reflect the electronic readout scheme of the ASICs.
The simulations described here have been used to optimize the design of readout ASICs which are currently in development at Brookhaven National Lab (Deptuch et al.). We plan to perform detailed comparison of simulated and measured data as soon as the first ASICs and detector/ASIC systems have been fabricated.

The results presented here can be compared with previously published 
simulated and experimental results.
\citet{BENOIT2009508} for example simulate a \SI{7.5}{mm} thick CZT detector with cross strip contacts at a pitch
of \SI{225}{\upmu m} and obtain good agreement between
simulated and experimental data. They report energy resolutions of approximately \SI{7}{keV} and \SI{14}{keV} FWHM for the \SI{59.5}{keV} $^{241}$Am line and \SI{122.1}{keV} $^{57}$Co line, respectively. The thick detectors suppress 
the low-energy tail evident in the energy spectra
presented in this paper.

\citet{2017SPIE10397E..02R} present measurement results obtained with the 
HEXITEC CZT/ASIC hybrids which feature pixels at a next-neighbor pitch of \SI{250}{\upmu m}
on \SI{1}{mm} thick CZT detectors, and compare the experimental results with simulated results.
Rather than modeling the spatially distributed energy depositions following the
impact of the primary photon and the evolution of the charge cloud in the detector as done
in our work, Ryan et al.\ used a phenomenological parameterization of the size of the charge clouds.
Using a 2-D Gaussian with a width $\sigma=\SI{23}{\upmu m}$, they were able to 
model the shape of the energy spectrum observed for a $\sim$\SI{20}{keV} X-ray beam from an X-ray gun.
Our approach is more general, and can be used over a wider range of energies.

The interested reader may consult  \citep{Iniewski:07,2009SPIE.7435E..03R,2011SPIE.8145E..07K,Yin:11,Yin:13,2014ITNS...61..154Y,2014NIMPA.767..218V,Monte:14,2018NIMPA.884..136O,Khalil:ay5516} 
for discussions of charge sharing in CZT detectors, and \citep[][]{2004ITNS...51.3098Z,BENOIT2009508,2008ITNS...55.1593D,2011NIMPA.654..233K,2012ITNS...59..236Z,2013NIMPA.708...88B,2014ITNS...61..154Y,2014JKPS...64.1336K,2015NIMPA.784..377W,2018SPIE10762E..0NC}
for discussion of CZT detectors of different thicknesses and pixel sizes. 

Detectors with pixels at \SI{150}{\upmu m} pixel pitch, as the ones discussed
in this paper, would be well suited for a {\it NuSTAR} follow-up 
mission with arcsecond angular resolution and a {\it NuSTAR}-like 
focal length of \SI{10}{m}. For these parameters, an HPD of \SI{5}{\arcsecond}
corresponds to a \SI{242}{\upmu m} focal spot. 
A pixel pitch of \SI{150}{\upmu m} would thus enable an oversampling factor of 1.6. 
A longer focal length of \SI{20}{m} and an HPD of \SI{15}{\arcsecond}, as proposed for 
HEX-P \citep{2018SPIE10699E..6MM}, the pixel diameter
would correspond to a spot diameter of \SI{1.5}{mm}, not necessitating 
sub-mm pixels.  

At energies of $\sim$\SI{70}{keV} and above, the \SI{1}{mm} thick detectors studied in this paper start to
become transparent, and the energy spectra start to develop an increasingly pronounced low-energy tail. The tail results from a combination of photons interacting close to the detector anodes so that electrons induce little charge before impinging on the anodes, and Compton scattered photons traveling beyond the next neighbor pixels or escaping the detector. Thicker detectors suppress both effects, but lead to longer drift
paths and thus to additional charge spreading and larger pixel multiplicities -- effects that adversely affect the energy resolution.
Covering the broad energy range from $\sim$\SI{2}{keV} to $\sim$\SI{160}{keV} or higher as envisioned for
{\it HEX-P} may require a layered detector, e.g. a front layer of thin ($\sim$\SI{1}{mm}) CZT detectors
recording lower-energy photons with excellent energy and spatial resolutions, followed
by a rear layer of thick ($\sim$\SI{1}{cm}) CZT detectors that catches the 
higher-energy photons with high efficiency and excellent resolutions.

Note that several alternatives to CZT detectors are currently being developed.
A possible alternative for finely-pixelated CZT detectors are Thallium Bromide (TlBr) detectors. The high atomic number of Thallium (81) make 
TlBr more efficient for photoelectric interactions than CZT, and 
\citet{2020FrP.....8...55K} report sub-1\% energy resolutions.
Germanium based Charge Coupled Devices (Ge-CCDs) may be another competitor, but still suffer from yield issues \citep{2019SPIE11118E..02L}.
We are currently evaluating the gamma-ray detectors made of tin absorbers and Transition Edge Sensors (TES) developed by the National Institute of Standards and Technology (NIST) \citep{2012RScI...83i3113B,2017ApPhL.111f2601M}.
The microcalorimeter detectors achieve superior energy resolutions 
(i.e. \SI{53}{eV} at \SI{97}{keV}) than solid state detectors.
For the time being, the drawback of the TES gamma-ray detectors 
are spatial resolutions on the order of \SI{\sim 1}{mm}.

Finely-pixelated CZT detectors as the ones discussed in this paper fill an important niche in the energy range of a few keV to $\sim$\SI{100}{keV} offering a unique combination 
of operation at room temperatures, 
high stopping power, high  photo-electric to Compton scattering 
cross sections, sub-mm spatial resolutions, and good energy resolutions.
%%%%%%%%%%%%%%%%%%%%%%%%%%%%%%%%%%%%%%%%%%%%%%%
\section*{Acknowledgements}
We thank Grzegorz Deptuch, Gabriella Carini, and Shaorui Li
for their work on the HEXID ASIC, as well as the McDonnell Center for the Space Sciences at Washington University in St.~Louis for its support. 
We thank Richard Bose and Andrew West for designing a HEXID readout system and HEXID photomasks. 
HK acknowledges NASA support under grants 80NSSC18K0264 and NNX16AC42G.

\bibliographystyle{elsarticle-harv}
\bibliography{hexPixel}
\end{document}